\documentclass[%
preprint,
amsmath,amssymb,
aps,
prfluids,
]{revtex4-2}

\usepackage[colorlinks,bookmarksopen,bookmarksnumbered,citecolor=red,urlcolor=red]{hyperref}

\usepackage{graphicx}
\usepackage{natbib}
\usepackage{color}
\usepackage{latexsym}
\usepackage{subfig}
\usepackage{natbib}
\usepackage{upgreek}
\usepackage{mathrsfs}
\usepackage{float}
\usepackage{caption}
\usepackage{soul}
\usepackage{textcomp}
\usepackage{stmaryrd}
\usepackage{amsmath}
\usepackage{lipsum} 
\usepackage{appendix}
\usepackage{xargs}   
\usepackage[normalem]{ulem}
\usepackage[pdftex,dvipsnames]{xcolor}  
\usepackage[colorinlistoftodos,prependcaption,textsize=small]{todonotes}
\usepackage{changes}
\usepackage{cancel}

\def\der#1#2{{\partial #1\over \partial #2}}

\newcommandx{\eyal}[2][1=]{\todo[linecolor=red,backgroundcolor=red!25,bordercolor=red,#1]{#2}}

\newcommandx{\anirban}[2][1=]{\todo[linecolor=blue,backgroundcolor=blue!25,bordercolor=blue,#1]{#2}}

\def\be{\begin{equation}}
\def\ee{\end{equation}}

\usepackage{cleveref}

\usepackage{color}
\definecolor{dgreen}{rgb}{0,0.5,0} 

\crefformat{section}{Sec. #2#1#3} 
\crefformat{subsection}{Sec. #2#1#3}
\crefformat{subsubsection}{Sec. #2#1#3}

\def\be{\begin{equation}}
\def\ee{\end{equation}}

\begin{document}

\preprint{APS/123-QED}

\title{Relating interfacial Rossby wave interaction in shear flows with Feynman's two-state coupled quantum system model for the Josephson junction} 

\author{Eyal Heifetz}
\affiliation{Department of Geophysics, Porter school of the Environment and Earth Sciences, Tel Aviv University,
Tel Aviv 69978, Israel.}
\email{eyalh@tauex.tau.ac.il, bratspiess@mail.tau.ac.il}

\author{Nimrod Bratspiess}
\affiliation{School of Physics and Astronomy, Tel Aviv University, Tel Aviv-Yafo 69978, Israel.}
\email{bratspiess@mail.tau.ac.il}


\author{Anirban Guha}
\affiliation{School of Science and Engineering Dundee, University of Dundee, Dundee DD1 4HN, United Kingdom}
\email{anirbanguha.ubc@gmail.com}

\author{Leo Maas}
\affiliation{Institute for Marine and Atmospheric research Utrecht, University of Utrecht, Utrecht 3584 CC, The Netherlands.}
\email{L.R.M.Maas@uu.nl}

\date{\today}

\begin{abstract} 

Here we show how Feynman's simplified model for the Josephson junction, as a macroscopic two-state coupled quantum system, has a one-to-one correspondence with the stable dynamics of two interfacial Rossby waves in piecewise linear shear flows.  
The conservation of electric charge and energy of the superconducting electron gas layers become respectively equivalent to the conservation of wave action and pseudoenergy of the Rossby waves. 
Quantum-like tunneling is enabled via action-at-a-distance between the two Rossby waves.  
Furthermore, the quantum-like phenomena of avoided crossing between eigenstates, described by the Klein-Gordon equation, is obtained as well in the classical shear flow system. In the latter, it results from the inherent difference in pseudoenergy between the in-phase and anti-phased normal modes of the interfacial waves. This provides an intuitive physical meaning to the role of the wavefunction's phase in the quantum system. A partial analog to the quantum collapse of the wavefunction is also obtained due to the existence of a separatrix between ``normal mode regions of influence'' on the phase plane, describing the system's dynamics. 
As for two-state quantum bits (qubits), the two-Rossby wave system solutions can be represented on a Bloch sphere, where the Hadamard gate transforms the two normal modes/eigenstates into an intuitive computational basis in which only one interface is occupied by a Rossby wave. Yet, it is a classical system which lacks exact analogs to collapse and entanglement, thus cannot be used for quantum computation, even in principle.

\end{abstract}

\maketitle


\section{Introduction}
\label{sec:1}

\begin{figure}
   \centering -
    \includegraphics[width=0.75\textwidth]{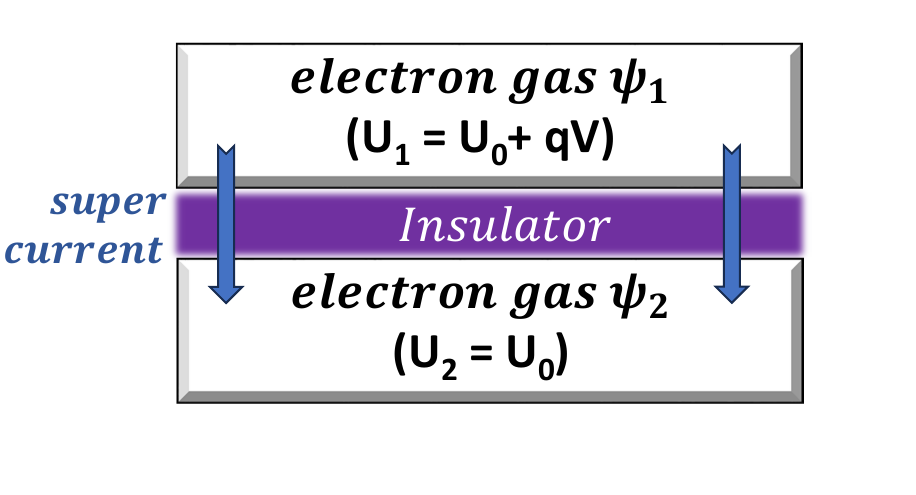}
   \caption{A Josephson junction device, detached from an electric circuit, is composed of two superconducting electron gas layers, separated by a thin insulator. The Cooper pairs in layer 2 are in the ground state energy $U_0$. Layer 1 was previously charged by voltage $V$, thus in energy state $U_0 +qV$ (where $q$ is the Cooper pair's electric charge). Each electron gas layer is described by a macroscopic wavefunction $\psi$, Eq.\,\eqref{Wavefunction12}. If the insulator is thin enough, a ``super current'' may flow via tunneling from one layer to the other.}
 \label{fig:fig1}
\end{figure}

In the very last chapter of {\it Feynman Lectures on Physics}  \cite{feynman1971feynman} (Vol. III, 21-9), Feynman suggested a simplified model for the Josephson junction. In this model, two layers (1 and 2) of superconducting electron gas (that is, a gas composed of quasi-particle Cooper pairs of electrons \cite{bardeen1957theory}, whose electrical resistance vanishes below a critical temperature) are separated by a thin insulating layer (Fig.~\ref{fig:fig1}). Since Cooper pairs behave as bosons, they tend to aggregate in the lowest possible energy quantum states, $U_{1,2}$, of each layer. 

Cooper pairs occupying the same quantum state are indistinguishable, thus for a very large number of pairs, the quantum probability density function $\rho$ for finding a single pair becomes their macroscopic density \footnote{This could be understood when adopting the logic behind the ensemble mean interpretation of the quantum probability density function. If we repeat again and again on an identical experiment, with a single quantum particle, then the  quantum probability density function converges to the density distribution function that is found when collecting the data of the particle positions in all the experiments. This can be considered as the rationale behind the relation of the fluid number
density of the Madelung fluid to the quantum probability density function \cite{heifetz2015toward}. Equivalently, if we consider at the same instant a very large number of indistinguishable quantum particles their quantum probability density function converges to the fluid number density of the superconducting gas}. Consequently, a macroscopic wavefunction can be assigned for each layer: 
\be
\label{Wavefunction12} 
\psi_{1,2} = \sqrt{\rho_{1,2}}e^{i\,\theta_{1,2}}\, ,
\ee
where it can be shown that in the absence of a magnetic vector potential, the gradients of the phases, $\nabla \theta_{1,2}$, are proportional to the momenta carried by the electron currents along each layer \cite{feynman1971feynman}. 

Feynman's simplified model assumes two homogeneous, non-magnetized, superconducting electron gas layers with zero current, thus 
$\psi_{1,2}$ are functions of time but not of space.
The electron gas in layer 1 was first charged by voltage $V$, making a difference in their lowest energies: $U_1 - U_2 = qV$ (where here $q$, is the electric charge of a Cooper pair, equal to twice the electron charge).
The model equations then read:  
\begin{subequations}
\label{FeynModel1} 
\begin{align}
i\hbar\dot{\psi}_1 =U_1 \psi_1 + K\psi_2\,  ,\\
i\hbar\dot{\psi}_2 =U_2 \psi_2 + K\psi_1\, .
\end{align}
\end{subequations}
Under perfect insulation between the two layers (when $K=0$), these are just the Schr\"{o}dinger equations for particles in the energy states $U_{1,2}$ ($\hbar$ is the reduced Planck constant). For non-zero values of the coupling constant $K$, the insulator is not perfect, hence quantum tunneling is possible between the layers, where $K$ generally decreases as the width of the insulator increases. 

Equation set  \eqref{FeynModel1} is an example of a two-state coupled quantum system. As will be shown in this paper, it also describes a seemingly unrelated system -- the stable interaction-at-a-distance between two interfacial Rossby waves in shear flows. This is quite intriguing as the latter system is purely classical. The main purposes of this paper are to understand how the physical processes in these two systems are related and what added value is obtained from a comparison between the two systems. 
In Section \ref{sec:2} we present the main physical properties of the superconducting system and then show, in Section \ref{sec:3}, how they reemerge in the context of the Rossby-wave system. Then in \ref{sec:4} we examine the quantum-like phenomena in the classical system and conclude, in Section \ref{sec:5}, with a discussion on the equivalency between the two systems.

\section{Properties of the Feynman model}
\label{sec:2}

\subsection{Representation as a dynamical system}
\label{subsec:2a}

Equation set  \eqref{FeynModel1} can be rewritten then as:
\begin{subequations}
\label{FeynModel2} 
\begin{align}
\dot{\psi}_1 =-i \hat{\omega}_1 \psi_1 - i \sigma \psi_2\,  ,\\
\dot{\psi}_2 =-i \hat{\omega}_2 \psi_2  - i \sigma\psi_1\, ,
\end{align}
\end{subequations}
where $\hat{\omega}_{1,2} = U_{1,2}/\hbar$, are effective frequencies associated with the energy states in the absence of interaction, and $\sigma = K/\hbar$, is the tunneling coupling coefficient, scaled by $\hbar$. 
Inserting the polar form of  Eq.\,\eqref{Wavefunction12} in Eq.\,\eqref{FeynModel2}, Feynman obtained dynamical equations for the densities and phases in the two layers. In terms of $\hat{\omega}_{1,2}$ and $\sigma$ we obtain:
\begin{subequations}\label{FeynModel3}
\begin{align}
& \hspace{-1.75cm} \dot{{\rho}_{1}} = -2\sigma \sqrt{\rho_{1}\rho_{2}}\sin{\delta} = -\dot{{\rho}_{2}}\, , \\
{\dot \theta}_{1}  = - \left({\hat \omega}_{1} + \sigma\sqrt{\rho_{2}\over \rho_{1}} \cos{\delta}\right)\, ; & \qquad {\dot \theta}_{2}  = - \left({\hat \omega}_{2} + \sigma\sqrt{\rho_{1}\over \rho_{2}} \cos{\delta}\right)\, , \\
& \hspace{-2cm} \dot{\delta} =  -\Delta{\hat \omega} +\sigma \left(\sqrt{\rho_{1}\over\rho_{2}} -\sqrt{\rho_{2}\over\rho_{1}}\right)\cos{\delta}\, ,
\end{align}
\end{subequations}
where $\delta = \theta_{1} - \theta_{2}$, $\Delta{\hat \omega} = {\hat \omega}_{1} - {\hat \omega}_{2}\,$ and $J = 2\sigma \sqrt{\rho_{1}\rho_{2}}|\sin{\delta}|$ is the magnitude of the ``super-current''
that may flow from one layer to the other by tunneling, through the insulator barrier
\footnote{Feynman pointed out that equation set  \eqref{FeynModel3} is incomplete when the Josephson junction is connected by wires to a battery of voltage $V$. Then the overall flow in the electric circuit keeps the charge density to be nearly constant throughout the circuit ($\rho_1 \approx \rho_2 = \rho_0$). Consequently, the magnitude of the super-current is $J_0 = 2\rho_0\sigma|\sin\delta|$, where the change in the phase $\dot{\delta} = -\Delta\hat{\omega}\,$, is only proportional to the voltage difference imposed by the battery. Here we refer only to the simpler setup in which a Josephson junction device stands alone, where the two electron gas layers are prepared a-priori to be in different energy states}.

\subsection{Hamiltonian and Hamilton equations}
\label{subsec:2b}

Equation (\ref{FeynModel3}a) indicates that the total charge, $\rho = \rho_1 + \rho_2$, is conserved. Equation set (\ref{FeynModel3}b) are scaled Hamilton-Jacobi density equations,  ${\dot \theta}_{i} = -{\cal H}_{i}$ \cite{heifetz2015toward}, where ${\cal H}_{i}$ is the Hamiltonian density (scaled by $\hbar$) of each layer $i=1,2$. Assuming that the volume of each layer is the same the  Hamiltonian per unit volume, $H = \rho_1{\cal H}_1 + \rho_2{\cal H}_2$, is conserved: 
\begin{equation}
\label{Hamiltonian}
H =  \sum_{i=1}^2  \rho_i{\cal H}_i  = -\sum_{i=1}^2 \rho_i{\dot \theta}_{i} =
\sum_{i=1}^2\rho_i{\hat \omega}_{i}  + 2\sigma\sqrt{\rho_1\rho_2}\cos{\delta}\, ,
\end{equation}
where substitution in Eq.\,\eqref{FeynModel3} verifies $\dot{H} = 0\,$.
Equivalently,  Eq.\,\eqref{FeynModel2} can be written in the matrix form:
\begin{align}\label{FeynModelMatrix}
i{\dot {\vec \Psi}} = {\hat {\bf  {\cal H}}} {\vec \Psi}\, , \hspace{0.2cm} 
\text{where} \hspace{0.2cm} {\vec \Psi} = \left(\begin{array}{c}
\psi_1\\
\psi_2
\end{array}\right)\,  \hspace{0.2cm} \& \hspace{0.3cm}
{\hat {\bf  {\cal H}}} =   \left(\begin{array}{cc}
{\hat \omega}_1  & \sigma\\
\sigma & {\hat \omega}_2
\end{array}\right) = {\hat {\bf  {\cal H}}}^T\, 
\end{align}
are the wavefunction vector in the Hilbert space and the Hermitian Hamiltonian density operator (superscript $(.)^T$ denotes the Hermitian transpose). 
The Hamiltonian in bra-ket notation (scaled by the volume) reads:
\begin{equation}\label{HamiltonBraKet}
H = \langle \Psi | {\hat {\bf  {\cal H}}} | \Psi \rangle \equiv
{\vec \Psi}^T  {\hat {\bf  {\cal H}}} {\vec \Psi} \, ,
\end{equation}
where it can be verified by direct substitution that the RHS of Eq.\,\eqref{Hamiltonian} and Eq.\,\eqref{HamiltonBraKet} are the same. 
The vanishing of the time derivative of $H$ in Eq.\,\eqref{HamiltonBraKet} is obtained directly as well when writing ${\dot {\vec \Psi}} = -i{\hat {\bf  {\cal H}}} {\vec \Psi}\,$ and ${\dot {\vec \Psi}^T} = i{\vec \Psi}^T{\hat {\bf  {\cal H}}}\,$.

Equations (\ref{FeynModel3}a,b) can  then be written in the compact form of the canonical Hamilton equations:
\begin{equation}\label{HamiltonEqs}
\dot{\rho}_i = \der{H}{\theta_i}\, , \qquad 
\dot{\theta}_i = -\der{H}{\rho_i}\, ,
\end{equation}
implying that in the 
four dimensional phase space: ${\bf x} \equiv (\rho_1, \rho_2, \theta_1, \theta_2)$,  the Hamiltonian $H$ serves as the streamfunction of the associated velocity field: 
${\bf u} \equiv ({\dot \rho_1}, {\dot\rho_2}, {\dot\theta_1}, {\dot \theta_2})$, so that ${\bf u}\cdot\nabla H = \{H,H\}  = 0$, where $\nabla$ is the four-component nabla operator in the phase space of coordinates ${\bf x}$. Furthermore, $\{f,H\} = \sum_{i=1}^2 \left( \der{H}{\theta_i}\der{f}{\rho_i} - \der{H}{\rho_i}\der{f}{\theta_i}\right) = {\bf u}\cdot\nabla f$ is the Poisson bracket of a general function $f({\bf x},t)$ with $H$. The phase space flow is volume preserving (i.e., divergence-free, 
$\nabla\cdot{\bf u} = 0$),  in agreement   with Liouville's theorem \cite{moretti2023analytical}.

\subsection{Eigenstates and avoided crossing}
\label{subsec:2b}

As will be shown here, the frequency eigenvalues of the two-level system of Eq.\,\eqref{FeynModel2} exhibit the phenomena of avoided crossing (e.g.~Ref.\,\cite{landau2013quantum} in the context of diatomic molecules),  i.e., the two frequency eigenvalues of the coupled system never equal to each other (their values `never cross'). This is valid even in the degenerated case where the two separated frequencies of the uncoupled system are equal (when 
$\hat{\omega}_1 = \hat{\omega}_2$).

The time independent Schr\"{o}dinger equation:
\begin{equation}\label{ISE}
\omega {\vec \Phi} = {\hat {\bf  {\cal H}}} {\vec \Phi}\, , 
\end{equation}
for ${\vec \Psi} = {\vec \Phi}e^{-i\omega t}$ (with the corresponding energy eigenvalues $E = \hbar \omega$), gives the frequency eigenvalues:
\begin{equation}\label{omega12}
\omega_{\pm} = \overline{\hat{\omega}} \pm \left[\left({1\over 2}\Delta \hat{\omega}\right)^2 +\sigma^2 \right]^{1/2}\, ,
\end{equation}
where $\overline{\hat{\omega}} = (\hat{\omega}_1 +\hat{\omega}_2)/2\,$.
As expected, in the absence of tunneling ($\sigma =0$) the two layers are uncoupled: $\omega_{+} = \hat{\omega}_{1}$, and $\omega_{-} = \hat{\omega}_{2}$, with the corresponding eigenvectors ${\vec \Phi}_+^T = (\sqrt{\rho_1}, 0)$; ${\vec \Phi}^T_- = (0, \sqrt{\rho_2})\,$. 

\begin{figure}
   \centering -
    \includegraphics[width=1.1\textwidth]{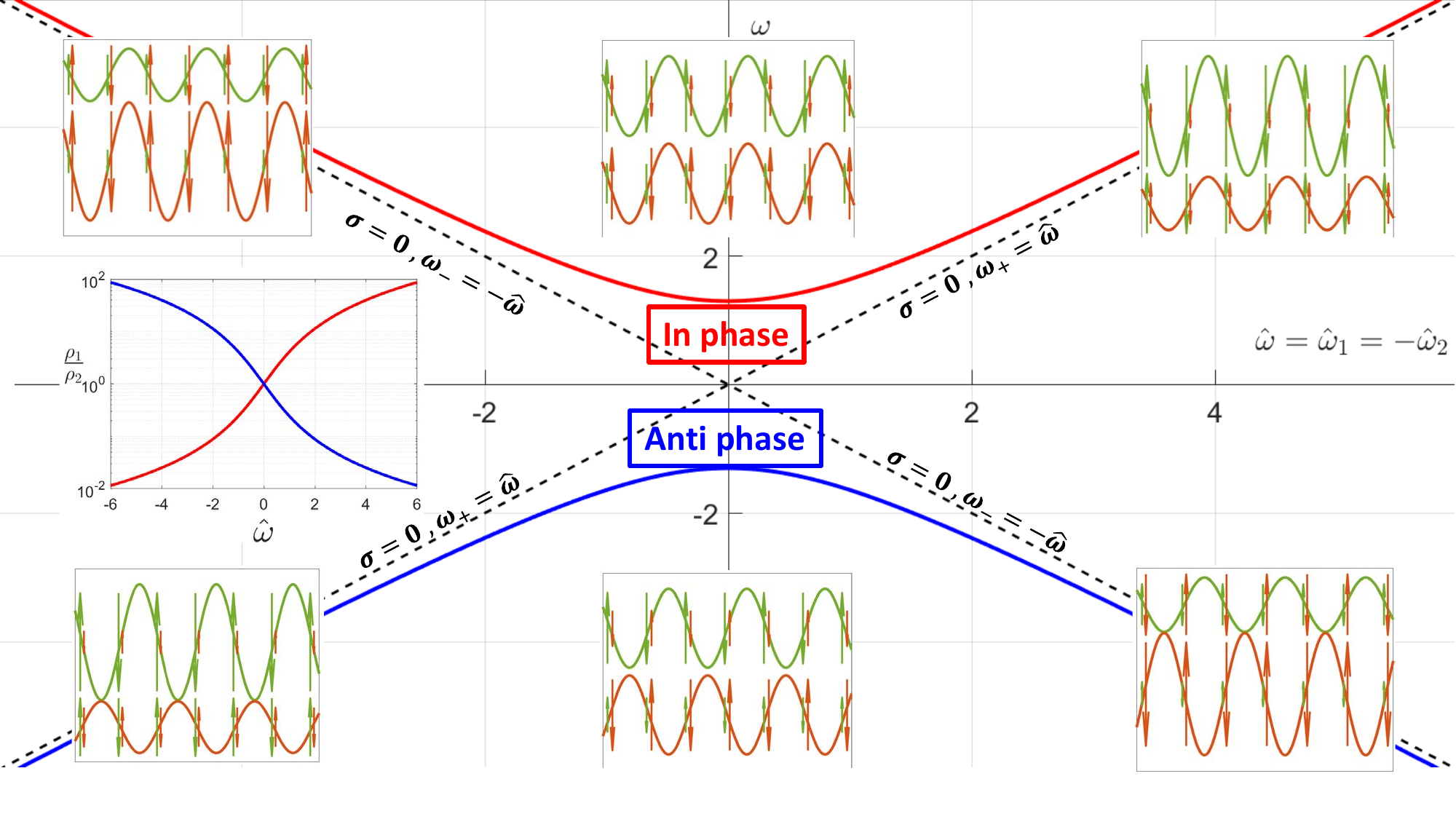}
   \caption{Eigenstate frequencies, $\omega_{\pm}$, satisfying the Klein-Gordon dispersion relation Eq.\,\eqref{Klein-Gordon} for tunneling coupling coefficient $\sigma = 1.5$. The upper (lower) hyperbola corresponds to eigenstate solutions in which the two layers wavefunctions are in (anti) phase. Dashed lines correspond to the uncoupled solutions ($\sigma = 0$) of each electron gas layer in isolation. The avoided crossing phenomena is evident as the two hyperbolas never cross, even on the ordinate where the frequencies of each layer in isolation are equal. The inset on the middle-left shows the eigenstate density ratio for the two hyperbolas, as obtained from Eq.\,\eqref{eigenstates}. The eigenstate structures of the Josephson junction system are visualized in the insets (with red and green waves and arrows) in terms of the normal modes of the equivalent Rossby waves system (see a detailed explanation for the interaction mechanism in subsections \ref{subsec:4a} and \ref{subsec:4b}).}
 \label{fig:fig2}
\end{figure}

Without loss of generality, we can move to the frame in which $\hat{\omega}_1  = -\hat{\omega}_2 \equiv \hat{\omega}$, so that $\overline{\hat{\omega}} =0$ and $\Delta \hat{\omega}/2 = \hat{\omega}$. Equation  \eqref{omega12}  then simplifies to the Klein-Gordon dispersion relation:
\begin{equation}\label{Klein-Gordon}
\omega_{\pm}  = \pm \left(\hat{\omega}^2 +\sigma^2 \right)^{1/2} \equiv \pm \Omega\, .
\end{equation}
In Fig.~\ref{fig:fig2}, we plot $\omega_{\pm}$ as a function of $\hat{\omega}$. 
For $\sigma=0$, the dashed straight lines in $\pm 45^{\circ}$ correspond to the solutions $\omega_{+} = \hat{\omega}_1$ and $\omega_{-} = \hat{\omega}_2$, respectively.
In the presence of tunneling, $\sigma>0$, 
the coupled solutions of the upper and lower hyperbolic branches never cross (even in the degenerate case  where $\hat{\omega}_1 = \hat{\omega}_2 =0$). 

The corresponding two orthonormal eigenstate vectors ${\vec \Phi}_{\pm}$, obtained from Eq.\,\eqref{ISE} are: 
\begin{equation}\label{eigenstates}
{\vec \Phi}_{+} = {1\over \sqrt{2}}
\left(\begin{array}{c}
\sqrt{\Omega + \hat{\omega}\over \Omega}
\\
\sqrt{\Omega - \hat{\omega}\over \Omega}
\end{array}\right)\, ; \qquad
{\vec \Phi}_{-} = {1\over \sqrt{2}}
\left(\begin{array}{c}
\sqrt{\Omega - \hat{\omega}\over \Omega}
\\
-\sqrt{\Omega + \hat{\omega}\over \Omega}
\end{array}\right)\, ,
\end{equation}
where the general solution of Eq.\,\eqref{FeynModel2} is superposition of the eigenstates:
\begin{equation}\label{generalsol}
{\vec \Psi}(t) = a_+ {\vec \Phi}_{+}e^{-i\omega_{+} t} +
a_- {\vec \Phi}_{-}e^{-i\omega_{-} t}\, 
\end{equation}
($a_{\pm} = A_{\pm}e^{i\alpha_{\pm}}$ are two complex amplitudes).
Since ${\vec \Phi}_{+}$ and ${\vec \Phi}_{-}$ are orthonormal,  
$a_{\pm} = \langle \Phi_{\pm}|\Psi(t=0)\rangle\,$.
Hence, the seemingly complicated nonlinear dynamics in Eq.\,\eqref{FeynModel3} can be projected into the superposition of the two eigenvectors, where each propagates with its constant eigenvalue  frequency. 
For example, starting from an initial condition of zero density in the lower layer: ${\vec \Psi}^T(t=0)= (1,0)^T$, yields 
$a_{\pm} = {\phi_1}_{\pm}$ (where ${\phi_1}_{\pm}$ corresponds to the first entries of the vectors ${\vec \Phi}_{\pm}$). 
Furthermore, for each pure eigenstate solution:
\begin{equation}\label{Pureeigenstates}
{\vec \Psi}_{\pm} = \left(\begin{array}{c}
\sqrt{\rho_1}e^{i\theta_1}\\
\sqrt{\rho_2}e^{i\theta_2}
\end{array}\right)_{\pm} = a_{\pm}
{\vec \Phi}_{\pm}e^{-i\omega_{\pm} t} = 
\left(\begin{array}{c}
\sqrt{\rho_1}e^{i(\theta^0_1 -\omega t)}\\
\sqrt{\rho_2}e^{i(\theta^0_2 -\omega t)}
\end{array}\right)_{\pm}\, 
\end{equation}
(here the zero superscript corresponds to the initial time $t=0$),
$\rho_{1,2}$, their ratio, 
$\rho_{1}/\rho_{2}$, as well as $\theta^0_{1,2}$ are all constant. Since $\theta_{1,2} = \theta^0_{1,2} - \omega t$, the phase difference $\delta = \theta_1 - \theta_2 =\theta^0_1 - \theta^0_2 \,$ is also constant, where $\dot{\theta}_1 = \dot{\theta}_2 = -\omega\,$.
As ${(\Omega \pm \hat{\omega})}$ are both positive (for non zero positive values of $\sigma$), Eq.\,\eqref{eigenstates} indicates that $\delta_+ = 0$ and $\delta_- = \pi$. Thus, the upper branch of the hyperbola corresponds to the cases where the upper and lower macroscopic wavefunctions are in phase, whereas the lower branch corresponds to the cases where they are anti-phased.  
The eigenstates energy then satisfy Eq.\,\eqref{Hamiltonian}:
\begin{equation}
\label{HamiltonianEigen}
H_{\pm} =  (\rho_1 - \rho_2)\hat{\omega} \pm 2\sigma\sqrt{\rho_1\rho_2} = \rho\, \omega_{\pm}\, .
\end{equation}
We recall that $H$ is scaled by $\hbar$, which is the elementary action unit of a quantum particle. Thus, when integrating $\rho \hbar$ over the volume of the two electron gas layers we obtain the total action of the Cooper pairs in the system. We will indeed see that in the Rossby wave system,  $\rho$ corresponds to their wave action. 

While the mathematical analysis shown above is straightforward, the physical interpretation of the dynamics is somewhat obscured. For the eigenstate solutions, the layers interact with each other via tunneling but no super-current flows from one layer to the other. The phase difference in the wavefunctions affects the charge density ratio between the layers and acts as well to synchronize the phase rate in the two layers. This is despite the fact that phase gradients are zero and, thus do not correspond to currents within the layers. 
As will be discussed in the following sections,
these issues become clear when implementing Feynman's model to describe the seemingly unrelated stable interaction-at-a-distance between interfacial Rossby waves in piecewise linear shear flows. 

\section{Feynman's model applied for interfacial Rossby waves}
\label{sec:3}

\subsection{Wave action and pseudoenergy conservation}
\label{subsec:3a}

For our purposes it is sufficient to consider inviscid, constant density, two-dimensional horizontal flows in the $x$--$y$ plane. The linearized dynamics of monochromatic small perturbations, with respect to a mean shear profile $\overline{u}(y)$ (denoted by overbars), satisfies the momentum, vorticity and continuity equations: 
\be
\label{EQsLinFlow}
{D {\bf u}\over Dt} = -\nabla p +\hat{\bf x}\,\overline{q}\,v\, ; \qquad {D q \over Dt} = -{\overline{q}}_{y}\,v \, ; \qquad \nabla \cdot {\bf u}= 0\, .
\ee
Here ${\bf u} = (u,v)$ is the perturbation velocity with respect to the mean flow $\overline{u}(y)$, $p$ is the pressure perturbation scaled by the constant density, and $q = \hat{\bf z}\cdot( \nabla\times {\bf u}) = (\der{v}{x} - \der{u}{y})$ is the vertical component of the vorticity perturbation (the hat superscript denotes a unit vector and  $\nabla$ is the horizontal nabla operator). The linearized material derivative is ${D \over Dt}  =  \der{}{t} + \overline{u}\der{}{x}$ and the mean flow vorticity and its gradient are respectively $\overline{q}(y) = -\der{\overline{u}}{y}$ and ${\overline{q}}_{y}  \equiv  \der{\overline{q}}{y} = -\der{^2\overline{u}}{y^2}$. 

We denote the cross-stream displacement (in the $y$ direction) by $\eta$ so that   $v = {D \eta \over Dt}$; the perturbation vorticity equation in Eq.\,\eqref{EQsLinFlow} then yields ${D\over Dt}(q + {\overline{q}}_{y}\eta) = 0$. We assume  
isovortical dynamics in which the vorticity perturbation results solely from the deformation of the basic state vorticity, $q = -{\overline{q}}_{y}\eta$, thus the vorticity perturbation can be non-zero only in regions where ${\overline{q}}_{y} \neq 0$. 
For a horizontal domain $(-L_x/2 < x < L_x/2\, ; -L_y/2 < y < L_y/2)$ with periodic stream-wise boundary conditions and vanishing cross-stream fluxes on the stream-wise boundaries, $v(y= \pm L_y/2)=0$, the linearized dynamics of Eq.\,\eqref{EQsLinFlow} conserves the two constants of motion - the pseudomomentum (and its associated wave action) and pseudoenergy \cite{heifetz2009canonical}. Define the stream-wise averaging as $\overline{(...)} \equiv {1\over L_x }\int_{-L_x/2}^{L_x/2}(...)dx$ and its integration in the cross-stream direction as 
$\langle ... \rangle \equiv \int_{-L_y/2}^{L_y/2} \overline{(...)} dy\,$, the pseudomomentum: 
\be
\label{PM}
P =  -{1\over 2}\biggl< {{q}^2 \over  {\overline{q}}_{y}} \biggr> =  -{1\over 2}\biggl<{{\overline{q}}_{y}\,{{\eta}^2}}  \biggr>\, ,
\ee
is the additional mean stream-wise momentum imposed by the perturbation on the flow.
If the mean vorticity gradient ${\overline{q}}_{y}$ does not change sign within the domain, this ensures stability as the vorticity perturbation cannot grow in time without violating the conservation of ${P}$
\footnote{The conservation of pseudomomentum can be therefore regarded as a generalization of Rayleigh inflection point criterion to ensure stability \cite{drazin2004hydrodynamic,heifetz2019normal}. As ${\overline{q}}_{y} = -\der{^2\overline{u}}{y^2}$, this implies that the basic state velocity profile does not posses an inflection point. Also, to avoid confusion, the middle term of Eq.\,\eqref{PM} does not imply that ${P}$ is singular in regions of constant shear where the mean vorticity gradient is zero. As indicated from the RHS term, it simply means that in these regions the cross-stream displacements do not generate vorticity perturbations and hence do not contribute to the pseudomomentum integral.}.
For a monochromatic perturbation of the form $f = \tilde{f}(y,t,k)e^{ikx}$, where $k$ is the stream-wise wavenumber (and $f$ is a general field), the wave action: \be
\label{Action0}
{\cal A} = {P\over k}\, 
\ee
is conserved as well. 
The second invariant is the pseudoenergy which is the additional kinetic energy imposed  by the perturbation on the base flow: 
\be
\label{PE}
E = {1\over 2}\biggl< {{\bf u}}^2 - {{\overline u}\over {\overline{q}}_{y}}{q}^2\biggr> = {1\over 2}\biggl< {{\bf u}}^2 - {{\overline u}\, {\overline{q}}_{y}}\, {\eta}^2\biggr>  \, ,
\ee
from which a generalization of the additional stability condition of Fj\o rtoft, $\mathrm{sgn}({\overline u}) = -\mathrm{sgn}({\overline{q}}_{y})$, is deduced \cite{drazin2004hydrodynamic,heifetz2019normal}.

From the incompressibility condition, the  perturbation streamfunction $\phi$ satisfies:
\be
\label{SF}
\nabla\cdot  {\bf u} = 0 \implies  {\bf u} = {\hat {\bf z}}\times \nabla \phi \implies q = \nabla^2 \phi  \, . 
\ee
The eddy kinetic energy integral can be written then as 
$K = {1\over 2}\bigl< {{\bf u}}^2 \bigr> = -{1\over 2}\bigl< q\,\phi\bigr>$ , 
which allows writing the pseudoenergy as:
\be
\label{PE2}
E = -{1\over 2}\biggl< q\left(\nabla^{-2} + {{\overline u}\over {\overline{q}}_{y}}\right)q\biggr> \, , 
\ee
where $\nabla^{-2}$ denotes the inverse horizontal Laplacian operator.  
Consider hereafter monochromatic wave dynamics, the RHS of Eq.\,\eqref{SF} yields:
\be
\label{SFK}
\tilde{q}(y,t,k) = \nabla_k^2\, \tilde{\phi} = \left(-k^2 +\der{^2}{y^2}\right)\tilde{\phi} \implies 
\tilde{\phi}(y,t,k) =   \nabla_k^{-2}\, \tilde{q} = 
\int_{-L_y/2}^{L_y/2}\tilde{q} (y',t,k) G_k(y,y')dy' \, ,
\ee
where $G(y,y')$ is the Green function, satisfying: 
$\nabla_k^2 G_k(y,y') = \delta(y-y')\,$ and $\delta$ is the Dirac delta function.  
For the purpose of the analysis here it is enough only to consider the open domain case: $L_y \rightarrow \infty$ for which $G_k(y,y') = -{e^{-k|y-y'|}/ 2k}$ \cite{heifetz2005relating}.

Writing the vorticity perturbation in terms of amplitude and phase: $\tilde{q} = Qe^{i\theta}$, when recalling that for two general monochromatic functions: $(f,g) =(\tilde{f}, \tilde{g})e^{ikx} =(Fe^{i\theta_f},Ge^{i\theta_g} )e^{ikx}$ we obtain 
$\overline{f\,g} = {1\over 2}\Re\{\tilde{f}\tilde{g}^*\} = {1\over 2}F G \cos{(\theta_f - \theta_g)}$ (where asterisk denotes complex conjugate), the wave-action and the pseudoenergy become \cite{heifetz2018generalized}:
\be
\label{Action}
{\cal A} = -\lim_{L_y \rightarrow \infty} \int_{-L_y/2}^{L_y/2} {Q^2 (y) \over {4k\overline{q}}_{y}(y)} dy\, ;
\ee
\be
\label{ExplicitPE}
E = \lim_{L_y \rightarrow \infty} \int_{-L_y/2}^{L_y/2}{Q(y)\over 4}\left[
\int_{-L_y/2}^{L_y/2} Q(y')\cos{(\theta(y) -\theta(y'))}    {e^{-k|y-y'|}\over 2k}dy'
-{\overline{u}(y)\, Q (y) \over {\overline{q}}_{y}(y)} \right] dy\, .
\ee
It will be shown that the wave-action $A$ plays the equivalent role of the charge density $\rho$ in the Josephson junction and the pseudoenergy $E$ the role of the Hamiltonian $H$.

\subsection{Single interfacial vorticity wave}
\label{subsec:3b}

\begin{figure}
   \centering -
    \includegraphics[width=1\textwidth]{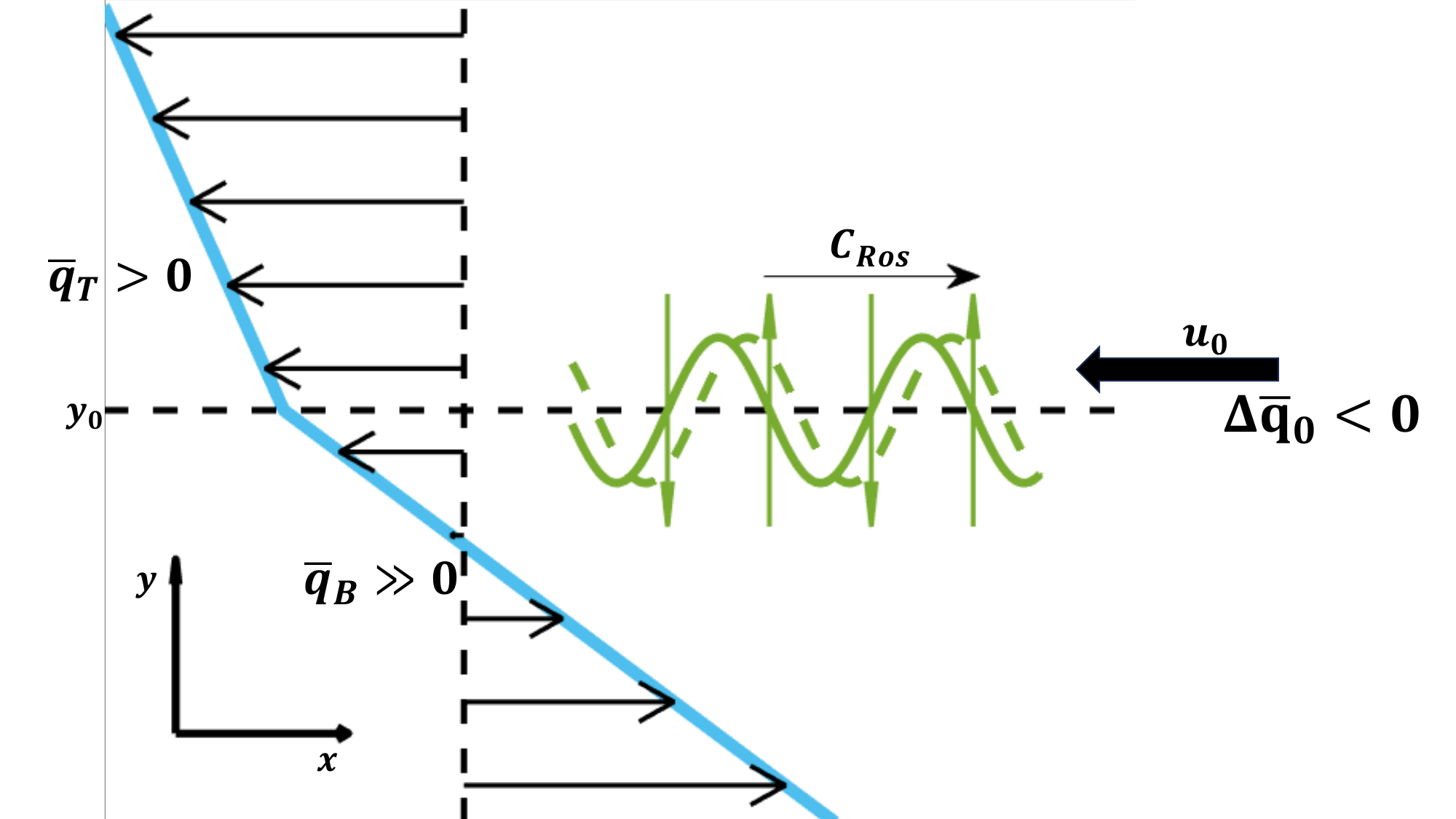}
   \caption{Schematic illustration of an interfacial counter propagating Rossby wave at $y=y_0$. The shear profile of Eq.\,\eqref{SingleProfile} has a negative  $\delta$-function value of mean vorticity gradient. Consequently, the Rossby wave acts to propagate to the right (leaving the lower mean vorticity to its left) counter the mean flow $\overline{u}_0$. The position of the wave after a short time interval is shown by the green dashed curve.}
 \label{fig:fig3}
\end{figure}

Consider first the piecewise linear shear flow profile (Fig.~\ref{fig:fig3}):
\begin{equation} \label{SingleProfile}
\overline{u}(y) = \overline{u}_0 -
\Bigg\{ \begin{array}{l@{\quad}l}
\overline{q}_T (y-y_0), & \textrm{$y \ge y_0$} \\
\overline{q}_B (y-y_0), & \textrm{$y \le y_0$} \\
\end{array}\,  ; \, \,
\overline{q}(y) =
\Bigg\{ \begin{array}{l@{\quad}l}
\overline{q}_T & \textrm{$y > y_0$} \\
\overline{q}_B  & \textrm{$y < y_0$} \\
\end{array}\,  ; \, \,\overline{q}_y = 
\Delta \overline{q}_0\, \delta(y-y_0)\, ,
\end{equation}
where $\overline{u}_0$ is negative, $(\overline{q}_{T}, \overline{q}_{B})$ are both positive constants and $\Delta \overline{q}_0 \equiv \overline{q}_T - \overline{q}_B\,$ is assumed negative.
The isovortical monochromatic vorticity perturbation has therefore a delta function structure: 
$\tilde{q} = -\tilde{\eta} \Delta \overline{q}_0\, \delta(y-y_0) \equiv \hat{q}_0(t)\delta(y-y_0)$. Substitute $\tilde{q}$ in the RHS equation of Eq.\,\eqref{SFK} we obtain:
\be
\label{SFKSingle}
\tilde{\phi}(y,t) = G_k(y,y_0)\hat{q}_0(t)  =  -
{e^{-k|y-y_0|}\over 2k}\hat{q}_0(t)\, .
\ee
While $\tilde{\phi}$ is continuous across the interface $y_0$,
$\der{\tilde{\phi}}{y}$ is discontinuous. Consequently $\tilde{v} = ik\tilde{\phi}$ is continuous, but $\tilde{u} = - \der{}{y}\tilde{\phi}$ flips sign across the interface. The latter yields an infinite perturbation shear, 
$- \der{u'}{y}$, across the interface, corresponding to the delta function structure of the vorticity perturbation. 

Substituting an interfacial vorticity wave perturbation $q'_0 = {\hat{Q}}_0\delta(y-y_0)e^{i(kx -\hat{\omega}_0 t)}$ (so that $\theta_0 =  -\hat{\omega}_0 t$) together with its associated streamfunction Eq.\,\eqref{SFKSingle} in the vorticity equation of Eq.\,\eqref{EQsLinFlow},  for profile Eq.\,\eqref{SingleProfile} we obtain at $y=y_0$:
\be
\label{thetadotzero}
\dot{\tilde{q}}_0 = -i\hat{\omega}_0\, \tilde{q}_0\, ; \qquad 
\dot{\theta}_0 = -\hat{\omega}_0\, , 
\ee
with the frequency:
\be
\label{FrequencySingle}
\hat{\omega}_0 = k \hat{c}_0 = k\left(\overline{u}_0 - 
{\Delta \overline{q}_0\over 2k}\right) = 
k\left(-|\overline{u}_0| + 
{|\Delta \overline{q}_0|\over 2k}\right) =
-k\left(|\overline{u}_0| +
|\Delta \overline{q}_0|  G^s \right),
\ee
where hereafter we use  $\dot{(...)} \equiv \der{}{t}(...)$ (in order to relate between the two common notations of temporal derivative in quantum and fluid mechanics), and $G^s \equiv -1/2k = G_k(y_0,y_0)$ is the self-induced Green function.
The propagation mechanism is of interfacial counter-propagating Rossby waves \cite{heifetz1999counter} (Fig.~\ref{fig:fig3}).  
The vorticity perturbation field at the interface induces a cross-stream velocity field, located a quarter of wavelength to its right. Since $\Delta\overline{q}_0$ is negative, the advection of the mean vorticity by the perturbation velocity across the interface results in fresh interfacial vorticity anomalies to the right. This acts  to translate the waves in the positive $x$ direction (the term $ 
{|\Delta \overline{q}_0|\over 2k}$ in Eq.\,\eqref{FrequencySingle}) counter the mean flow $-|\overline{u}_0|$. As the wavenumber $k$ is positive definite, the dispersion relation in Eq.\,\eqref{FrequencySingle} indicates that the sign of $\hat{\omega}_0$ is positive (negative) for $k < k_c$ ($k > k_c$), where $k_c \equiv 0.5{|\Delta \overline{q}_0|/ |\overline{u}_0|}$.
Hence, for long enough wavelength the waves overcome the mean flow Doppler shift and propagate to the right, when viewed from a frame of rest, whereas waves with smaller wavelengths than $2\pi/k_c$ are drifted to the left in the direction of the interfacial mean flow $\overline{u}_0$. Substituting Eq.\,\eqref{thetadotzero} and Eq.\,\eqref{FrequencySingle} in Eq.\,\eqref{Action} and Eq.\,\eqref{ExplicitPE} we obtain the single interface action-angle relations:
\be
\label{SingleActionAngle}
E_0 = - {\cal A}_0\dot{\theta}_0 = {\cal A}_0\hat{\omega}_0\, ; \qquad
{\cal A}_0 =  {\hat{Q}_0^2 \over 4k|\Delta\overline{q}_0|}\, .
\ee

\subsection{Two interfacial vorticity waves}
\label{subsec:3c}

\begin{figure}
   \centering -
    \includegraphics[width=1\textwidth]{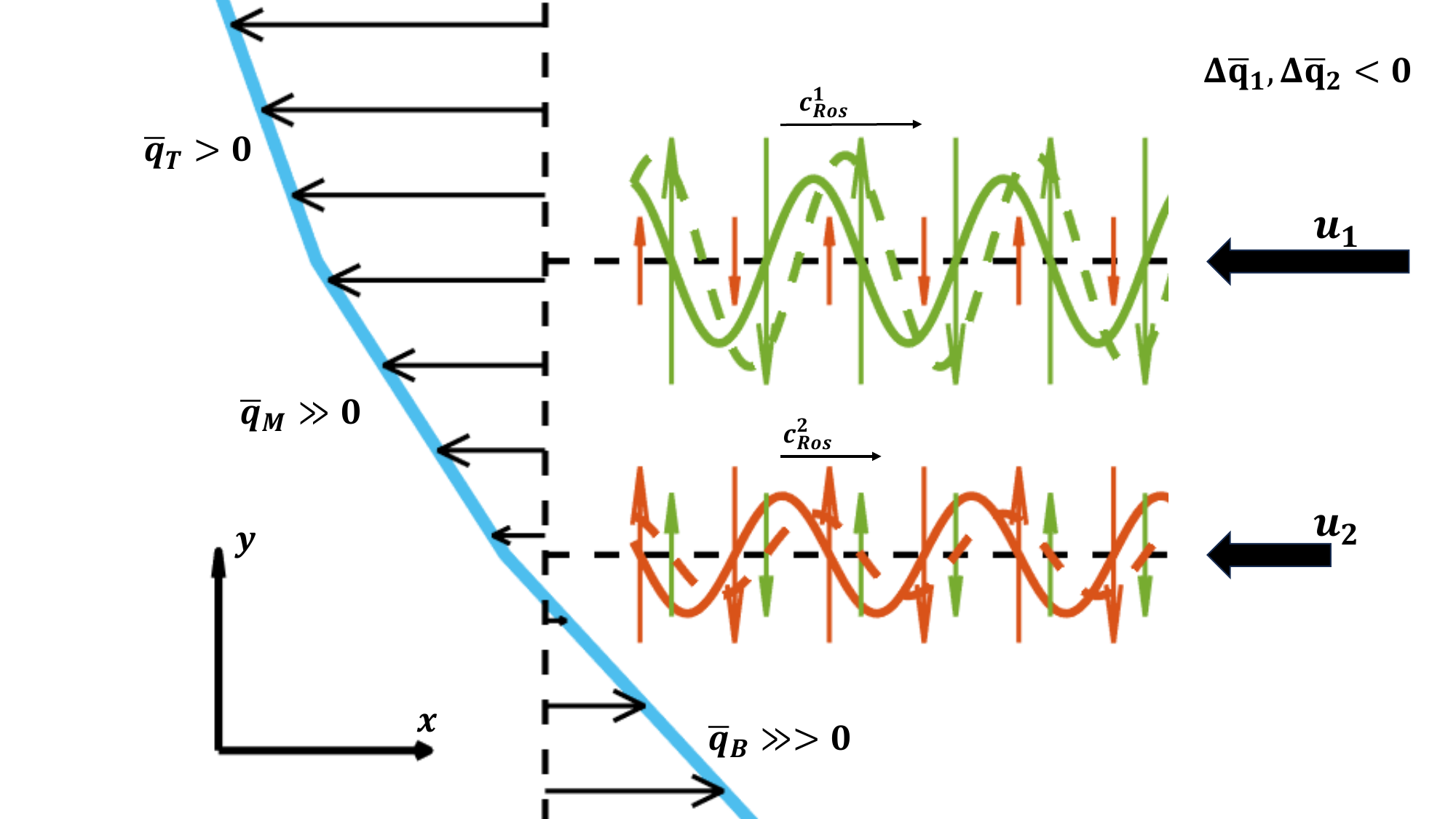}
   \caption{Schematic illustration of two interfacial counter propagating Rossby waves on the shear profile Eq.\,\eqref{TwoShear}. As in Fig.~\ref{fig:fig3}, each wave acts to propagate to the right counter the local mean flow at its interface. The waves interact at a distance by inducing their cross-stream velocity which decays exponentially with distance from their home bases. In this example the upper wave's amplitude is larger than the lower one. Consequently, the influence of the upper wave on the lower is larger than vice-versa. The phase difference between the waves at this snapshot is $-\pi/2 < \delta < 0$. As a result both waves amplify each others' rightward propagating speed, but have opposite effect on the amplitudes  - the lower (upper) wave increases (decreases) the upper (lower) wave's amplitude.}
 \label{fig:fig4}
\end{figure}

Consider now the shear flow profile (Fig.~\ref{fig:fig4}):
\begin{equation} \label{TwoShear}
\overline{u}(y) = 
\begin{cases}
\overline{u}_1 - \overline{q}_T (y-y_1), & \textrm{$y \ge y_1$} \\
\overline{u}_2 - \overline{q}_M (y-y_2), & \textrm{$y_2 \le y \le y_1$} \\
\overline{u}_2 - \overline{q}_B (y-y_2), & \textrm{$y \le y_2$} \\
\end{cases}\, ; \qquad
\overline{q}(y) =
\begin{cases}
\overline{q}_T, & \textrm{$y > y_1$} \\
\overline{q}_M, & \textrm{$y_2 < y < y_1$} \\
\overline{q}_B, & \textrm{$y < y_2$} \\
\end{cases}\,  ,
\end{equation}
so that 
\begin{equation} 
\label{TwoVortJumps}
\overline{q}_y = 
\Delta \overline{q}_1\, \delta(y-y_1) +
\Delta \overline{q}_2\, \delta(y-y_2) \, ,
\end{equation}
where $\overline{u}_1 < \overline{u}_2 < 0$ and  $\overline{q}_{B} > \overline{q}_{M} > \overline{q}_{T} >0$ $\implies$  $\Delta \overline{q}_1 \equiv \overline{q}_T - \overline{q}_M < 0$ and $\Delta \overline{q}_2 \equiv \overline{q}_M - \overline{q}_B < 0$.
Denote the distance between the interface by $Y \equiv y_1 - y_2$, Eq.\,\eqref{TwoShear} implies that $\overline{q}_{M} = -(\overline{u}_1 - \overline{u}_2)/Y\,$.

For isovortical dynamics, 
$\tilde{q} = \hat{q}_1(t)\delta(y-y_1) +
\hat{q}_2(t)\delta(y-y_2)$, with the corresponding streamfunction
$\tilde{\phi} =  -{1\over 2k}\left(\hat{q}_1 e^{-k|y-y_1|} +
\hat{q}_2 e^{-k|y-y_2|}\right)$. Substitute $\tilde{q}$ and $\tilde{\phi}$ in the vorticity equation of Eq.\,\eqref{EQsLinFlow} for $y=(y_1, y_2)$ (and recall that $\tilde{v} =  ik\tilde{\phi}$), we obtain:
\begin{subequations}
\label{TowWavesVort} 
\begin{align}
\dot{\hat{q}}_1 =-i \hat{\omega}_1 \hat{q}_1 - i \sigma_1 \hat{q}_2\,  ,\\
\dot{\hat{q}}_2 =-i \hat{\omega}_2 \hat{q}_2  - i \sigma_2 \hat{q}_1\, ,
\end{align}
\end{subequations}
with the wave frequencies, at the absence of interaction: 
\be
\label{omega12Rossby}
\hat{\omega}_{1,2} = k \hat{c}_{1,2} = k\left(-|\overline{u}| + 
{|\Delta \overline{q}|\over 2k}\right)_{1,2} = 
-k\left(|\overline{u}| +
|\Delta \overline{q}|  G^s \right)_{1,2}\, ,
\ee
and the interaction coefficients: 
\be
\label{sigma12}
\sigma_{1,2} = {e^{-kY}\over 2}|\Delta \overline{q}|_{1,2} = 
-k|\Delta \overline{q}|_{1,2}  G^i \, ,
\ee
where we denote $(G^s, G^i) \equiv 
-(1, e^{-kY})/2k\,$ as the self and induced values of the Green function of each wave on itself and on the opposed one.
Writing $\hat{q}_{1,2} = \hat{Q}_{1,2} e^{i\theta_{1,2}}$, the wave action of the perturbation can be obtained from Eq.\,\eqref{Action} to be:
\be
\label{Action12}
{\cal A} =  {\hat{Q}_1^2 \over 4k|\Delta\overline{q}_1|} +
{\hat{Q}_2^2 \over 4k|\Delta\overline{q}_2|} = {\cal A}_1 +{\cal A}_2\, .
\ee
Now, if we choose to define a ``wavefunction'' for the interfacial Rossby waves:
\be
\label{RossbyWF}
\psi^{Ros}_{1,2} \equiv\sqrt{{\cal A}_{1,2}}e^{i\,\theta_{1,2}} \qquad \implies \qquad \hat{q}_{1,2}  = 2\sqrt{k|\Delta\overline{q}|_{1,2}}\,  \psi^{Ros}_{1,2}
\ee
and substitute back in Eq.\,\eqref{TowWavesVort} we obtain:
\begin{subequations}
\label{RossbyWFl2} 
\begin{align}
\dot{\psi}^{Ros}_1 =-i \hat{\omega}_1 \psi^{Ros}_1 - i \sigma \psi^{Ros}_2\,  ,\\
\dot{\psi}^{Ros}_2 =-i \hat{\omega}_2 \psi^{Ros}_2  - i \sigma\psi^{Ros}_1\, ,
\end{align}
\end{subequations}
with the scaled interaction coefficient $\sigma  = \sqrt{|\Delta\overline{q}_1||\Delta\overline{q}_2|}\, {e^{-kY}/ 2}\, $.
Hence, assigning $\psi^{Ros}_{1,2} \mapsto \psi_{1,2}$ and ${\cal A}_{1,2} \mapsto\rho_{1,2}$ we obtain Eq.\,\eqref{FeynModel2} and thus Eq.\,\eqref{FeynModel3}. 

Using Eq.\,\eqref{ExplicitPE}, it is shown in Appendix~\ref{Appendix-B} that when assigning  $E \mapsto H$, Eq.\,\eqref{Hamiltonian} holds as well. 
Consequently, equations (\ref{FeynModelMatrix}-\ref{Pureeigenstates}) apply as well for the two Rossby wave system, thus completing the mathematical equivalence between the quantum and the classical systems.  

\section{Quantum-like effects in the Rossby wave system}
\label{sec:4}

\subsection{Interaction mechanism as tunneling}
\label{subsec:4a}

In Fig.~\ref{fig:fig4} we sketch an example for the interaction between the two interfacial waves. 
Since at each interface $\Delta \overline{q}$ is negative, the cross-stream displacement $\eta\,$, the vorticity field $q\,$, and the wavefunction $\psi^{Ros}$, are all in phase. At each interface the cross-stream velocity field is composed of two contributions: ``self'' (s), from the home-base wave and ``induced'' (i) from the remote wave. According to Eq.\,\eqref{SFK} at each interface $\tilde{v} = \tilde{v}^s + \tilde{v}^i  = i k(G^s \hat{q}^s + G^i \hat{q}^i)=-0.5\,i\,(\hat{q}^s + e^{-kY}\hat{q}^i)$.
Hence, $\tilde{v}^{s,i}$ is positioned a quarter of wavelength ahead of $\hat{q}^{s,i}$ where the action-at-a-distance interaction decays exponentially with the wavenumber, scaled by the distance between the interfaces. $\tilde{v}^i$ therefore accounts to the analog effect of tunneling in this classical system \cite{harnik2007relating}. More analysis on its nature is provided in Appendix~\ref{Appendix-A}.

In Fig.~\ref{fig:fig5} we show representative snapshots of the interaction according to their phase difference $\delta$. When $\delta = \pi/2$ (Fig.~\ref{fig:fig5}a) the upper wave reinforces the amplitude of the lower wave by inducing a velocity field which is in phase with the displacement of the upper wave. In contrast, the lower wave diminishes the amplitude of the upper wave as the induced velocity is in anti phase with the lower wave displacement. When $\delta = -\pi/2$ (Fig.~\ref{fig:fig5}b) the opposite scenario occurs. These are instantaneous snapshots as both the amplitude ratio $\rho_1/\rho_2$ (eq. \ref{FeynModel3}a) and the phase difference $\delta$ (eq.  \ref{FeynModel3}c) are continuously changing (unless $\Delta \hat{\omega} = 0$). Hence such phase configurations cannot describe eigenstates (in the quantum system), or equivalently normal modes (in the fluid system). We see that the waves cannot mutually amplify each other when the sign of their home base mean vorticity gradients is the same 
\footnote{This stands in contrast with the instability mechanism described in \cite{heifetz2019normal} where the mean vorticity gradients are of opposite signs. Consequently $\sigma_1 = -\sigma_2 = \sigma$, the propagator $\hat{\cal H}$ is non-Hermitian and both the pseudomomentum and the pseudoenergy integrals are zero for the unstable normal mode solutions.}.  

When the waves are in phase ($\delta = 0$), Fig.~\ref{fig:fig5}c, the self and the induced velocities are superposed, thus the waves help each other to propagate to the right with respect to  their local mean flow $\overline{u}$, (eq. \ref{FeynModel3}b). In contrast, in anti-phase ($\delta = \pi$), Fig.~\ref{fig:fig5}d, the induced velocity acts against the self velocity, thus slowing the rightward propagation tendency of each wave in isolation. Furthermore, if ${\hat{Q}_i / \hat{Q}_s} > e^{kY}$, the induced velocity overpowers the self one, forcing the wave to propagate to the left with respect to its local mean flow. These phase and anti-phased configurations may form the normal modes when $\dot{\delta} = 0$ (eq. \ref{FeynModel3}c), as discussed next.  

\subsection{Mechanistic interpretation for the normal modes and avoided crossing}
\label{subsec:4b}

\begin{figure}
   \centering     \includegraphics[width=1\textwidth]{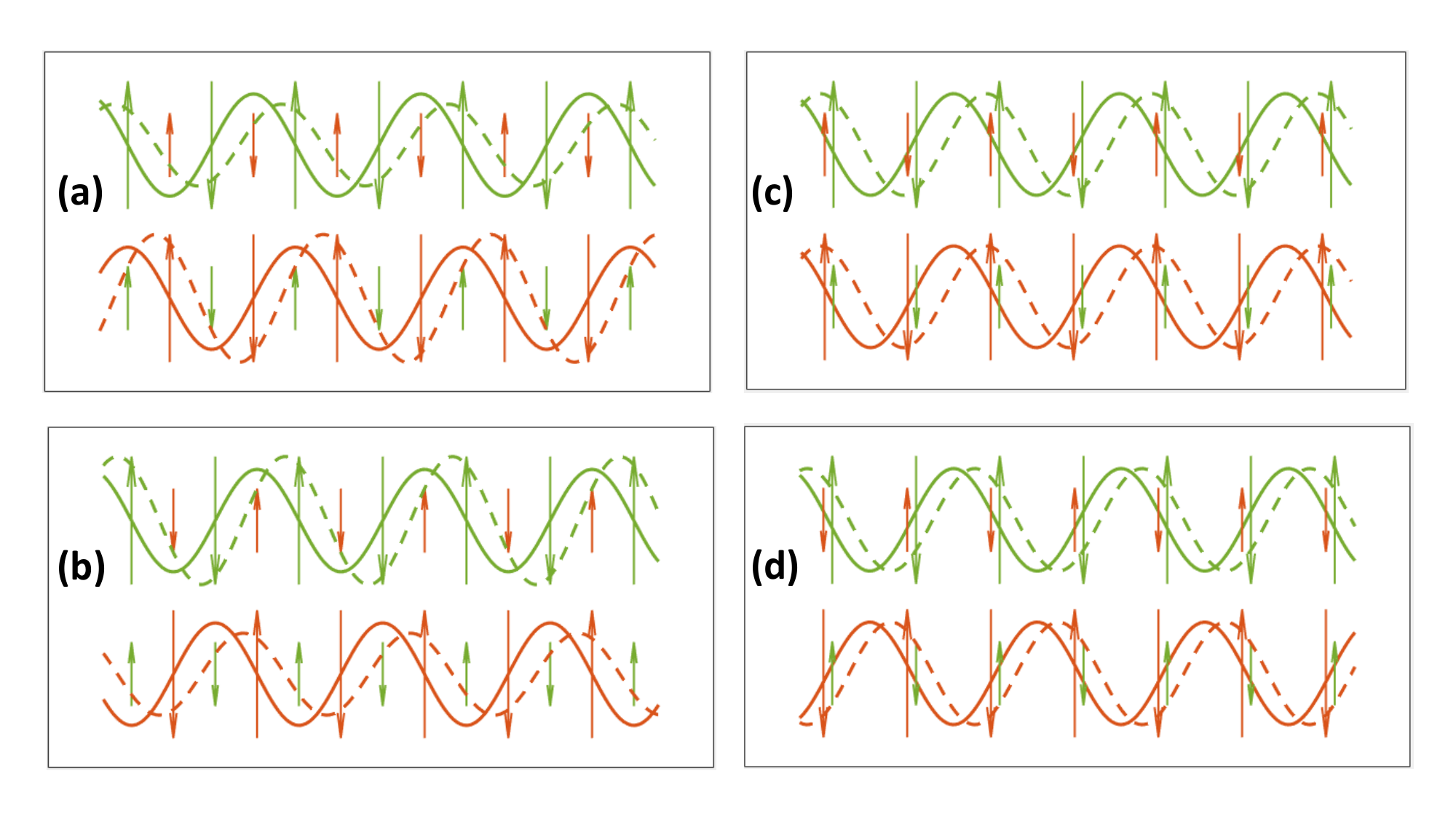}
   \caption{Representative snapshots of the waves action-at-a-distance interaction according to their phase difference $\delta =({\pi\over 2}, -{\pi\over 2},0,\pi)$ in (a,b,c,d), respectively. When $\delta = \pi/2$ the upper wave reinforces the amplitude of the lower wave while the lower wave diminishes the amplitude of the upper wave. When $\delta = -\pi/2$ the opposite occurs. When the waves are in phase ($\delta = 0$), they help each other to propagate to the right, while when in anti-phase ($\delta = \pi$), they slow each other.}
 \label{fig:fig5}
\end{figure}

As pointed out in the previous section, neutral normal modes in the fluid system (which are equivalent to the eigenstates Eq.\,\eqref{eigenstates} in the quantum system) can be obtained when the waves' amplitudes are either in phase ($\delta = 0$) or in anti-phase ($\delta = \pi$), (eq.  \ref{FeynModel3}a).
We wish to provide a mechanistic interpretation of the structure of these eigenstates/normal modes from the Rossby wave interaction perspective.  Recall that due to the negative sign of $\Delta \overline{q}$, at the two interfaces, each Rossby wave acts to propagate to the right, however, the local mean flow at each interface adds a Doppler shift toward the left direction. When the waves are in phase, they reinforce each other's tendency to propagate to the right and when they are in anti-phase, they hinder their rightward propagation. 

Consider first the case when $\hat{\omega} > 0$, that is in isolation, viewed from a frame of rest, the upper wave propagates to the right ($\omega_1 = \hat{\omega}$) and the lower to the left ($\omega_2 = -\hat{\omega}$). 
In the upper right hyperbola of Fig.~\ref{fig:fig2} the waves are in phase, where the upper wave's amplitude is larger than the lower one. Consequently, they help each other to propagate to the right but the upper wave provides more help to the lower one than vice-versa. The upper wave's amplitude should be large enough so that the velocity field it induces on the lower wave (although attenuated by $e^{-kY}$) will help the latter to overcome the leftward mean flow and consequently propagate  to the right. 
The lower wave helps (albeit, a little) the upper wave to propagate more to the right, thus their joint (eigen) frequency is larger than the frequency of the upper wave in isolation. 

In the lower right branch of the hyperbola the waves are anti-phased when the amplitude of the lower wave is larger than  the upper wave. Consequently, the velocity that the lower wave induces on the upper, opposes and overpowers the latter (despite the exponential attenuation), making the upper wave to propagate to the left. By the same time, the upper wave reduces by little the tendency of the lower wave to propagate to the right, thus consequently their joint (eigen) frequency is more negative than the frequency of the lower wave in isolation. 
Following the same logic, in the left side of Fig.~\ref{fig:fig2} (when $\omega_2 = -\omega_1  = -\hat{\omega} < 0$), in the upper left of the hyperbolas the waves are in phase, where the amplitude of the lower wave is larger than the upper. In the lower left part, the waves are anti-phased and the amplitude of the upper wave is larger than the lower. 

The avoided crossing phenomena is easily explained in terms of the wave interaction mechanism. In the degenerate case when each wave in isolation has the same frequency ($\omega_1 = \omega_2  = 0$) they still interact. When they are in phase and have the same amplitude they equally help each other to propagate, thus their joint eigenstate frequency is positive. When they are in anti-phase, with equal amplitudes, they equally hinder each other's propagation to the right and consequently their joint frequency becomes negative.

\subsection{Qubit-like representation on a Bloch sphere}
\label{subsec:4c}

\begin{figure}
   \centering    \includegraphics[width=1.3\textwidth]{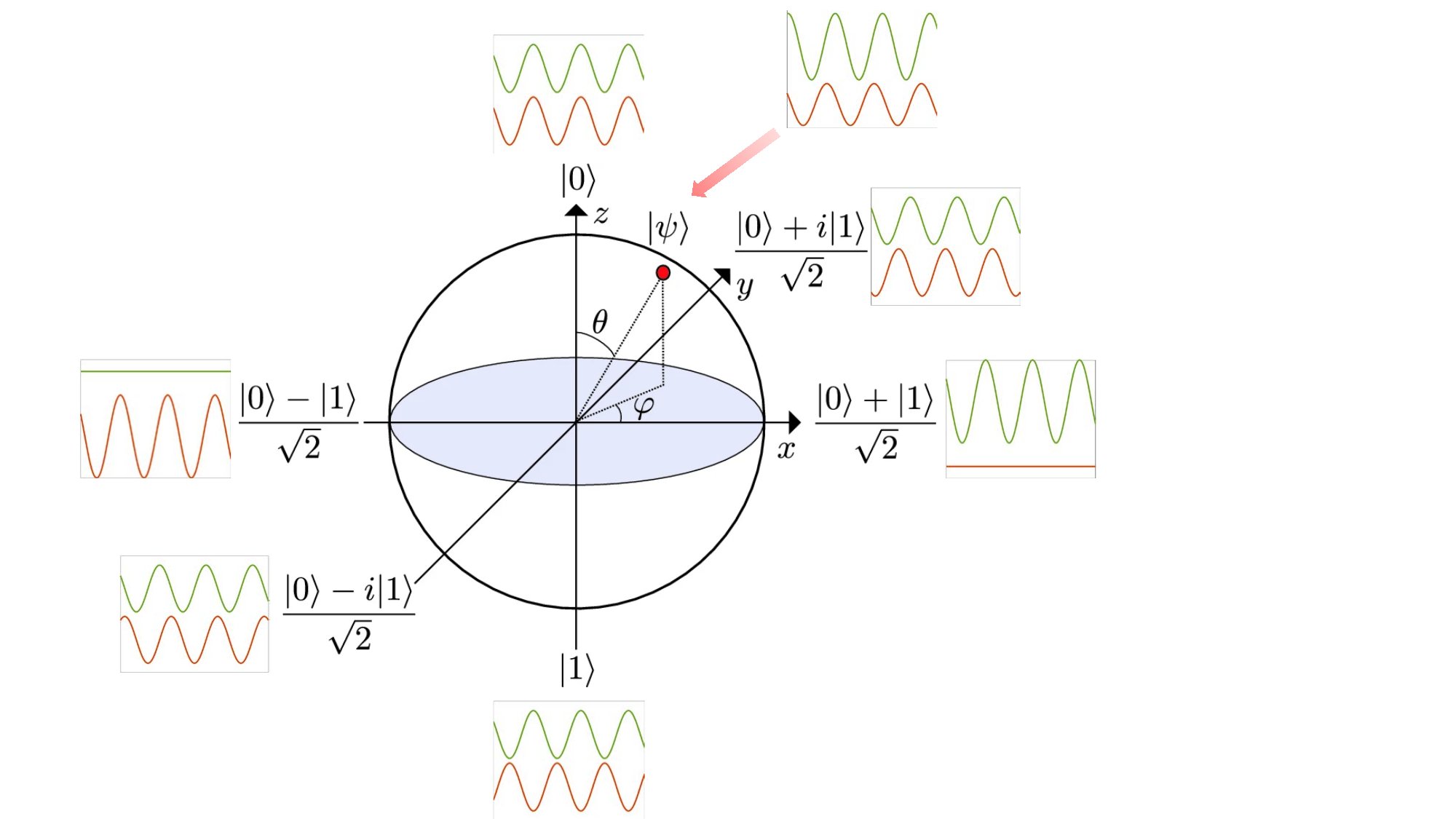}
   \caption{Bloch sphere representation adopted from Ref.\,\cite{kockum2019quantum}. Each quantum state $|\Psi \bigr>$ satisfying  Eq.\,\eqref{BlochSphere2} is a point on the sphere envelope, where each pair of opposite unit vectors form an orthonormal computational basis. 
   The two eigenstates $(|0 \bigr>, |1 \bigr>)$ are located on the $z$ axis where the $(|+ \bigr>, |- \bigr>)$ basis, obtained from applying the Hadamard gate on the eigenstates, are located on the $x$ axis. The general structure of $|\Psi \bigr>$ can be visualized in terms of snapshots of the two Rossby waves (here presented for the avoided crossing point of $\Delta \omega = 0$). The two eigenstates are Rossby waves in phase and anti-phase. The $(|+ \bigr>, |- \bigr>)$ basis represents the setups in which only one wave exists on each interface. The $\pm \pi/2$ out of phase setups of \ref{fig:fig5}(a,b) are obtained for the computational basis $\left(|0\bigr> \mp i |1\bigr>\right)/\sqrt{2}$.}
 \label{fig:fig6}
\end{figure}


Although the Rossby wave system is classical it shares the same dynamical equations for a two-state coupled quantum system, where the latter can be seen as a quantum bit (qubit) device for quantum computations \cite{mermin2007quantum}.
In this section, we apply quantum computation notation to describe the Rossby wave system. We denote the in-phase eigenstate ${\vec \Phi}_{+}$ in Eq.\,\eqref{eigenstates} as $|0\bigr>$, and the anti-phase eigenvector ${\vec \Phi}_{-}$ as $|1\bigr>$. Then, their superposition Eq.\,\eqref{generalsol} can be rewritten for $t=0$ as: 
\begin{equation}
\label{BlochSphere1}
|\Psi \bigr> = e^{i\alpha_+}\left[A_+ |0\bigr> + 
A_-e^{i(\alpha_- -\alpha_+)}|1\bigr> \right]\, . 
\end{equation}
As $|0\bigr>$ and $|1\bigr>$ are orthonormal, then for normalized $|\Psi \bigr>$, $A^2_+ + A^2_- = 1$. Hence, we can assign an angle $\theta$ so that $A_+ = \cos{\theta \over 2}$ and $A_- = \sin{\theta \over 2}$. Denoting the eigenstate phase difference as $\phi \equiv (\alpha_- -\alpha_+)$ and choosing $\alpha_+ = 0$ (without loss of generality), the superposition of  Eq.\,\eqref{BlochSphere1} reads:
\begin{equation}
\label{BlochSphere2}
|\Psi \bigr> = \cos{\theta \over 2}|0\bigr> + e^{i\phi}
\sin{\theta \over 2}|1\bigr>\, , 
\end{equation}
which can be represented on the Bloch sphere of unit radius \cite{mermin2007quantum}, as illustrated in Fig.~\ref{fig:fig6}.
Motion on the sphere envelope is obtained by applying a sequence of unitary transformations, where each pair of opposite unit vectors forms an orthonormal computational basis. For instance, the useful action of the Hadamard gate on the eigenstates
\begin{equation}
\label{Hadamard}
{1\over \sqrt{2}}
\begin{bmatrix}
1 & 1 \\
1 & -1 
\end{bmatrix}
\begin{bmatrix}
|0\bigr> \\
|1\bigr>
\end{bmatrix}
={1\over \sqrt{2}}
\begin{bmatrix}
|0\bigr> + |1\bigr>\\
|0\bigr> - |1\bigr>
\end{bmatrix} \equiv 
\begin{bmatrix}
|+ \bigr> \\
|-\bigr>
\end{bmatrix}\, 
\end{equation}
transforms $|0\bigr> \mapsto (|0\bigr> + |1\bigr>)/\sqrt{2}$ (rotates a unit vector in the positive $z$ coordinate into the positive $x$ coordinate) and $|1\bigr> \mapsto (|0\bigr> - |1\bigr>)/\sqrt{2}$ (rotates a unit vector in the negative $z$ coordinate into the negative $x$ coordinate). In other words, this transforms the orthonormal eigenstate computational basis $(|0\bigr>, |1\bigr>)$ from the $z$ coordinate into the orthonormal computational basis $(|+\bigr>, |-\bigr>)$ on the $x$ coordinate.

For the avoided crossing point of $\Delta\hat{\omega} = 0$, these two central computational basis (Fig.~\ref{fig:fig6}) become:
\begin{equation}
\label{basis}
|0\bigr> = {1\over \sqrt{2}}
\begin{bmatrix}
1   \\
1   
\end{bmatrix}; \qquad
|1\bigr> = {1\over \sqrt{2}}
\begin{bmatrix}
 1  \\
-1   
\end{bmatrix}, \qquad
|+\bigr> = 
\begin{bmatrix}
1   \\
0   
\end{bmatrix}; \qquad
|-\bigr> = 
\begin{bmatrix}
0   \\
1   
\end{bmatrix}\, .
\end{equation}
Thus,  the 
$(|+\bigr>, |-\bigr>)$ basis represents the two cases where only one wave at each interface exists (equivalent to that only one layer is filled with an electron gas and the other is empty). The assignment of the 
Rossby wave  Eq.\,\eqref{RossbyWF} to the quantum wavefunction
provides as well simple intuitive meaning for other computational basis. For instance, the application of the $S$ gate \cite{mermin2007quantum} rotates the eigenstates into $\left(|0\bigr> \pm i |1\bigr>\right)/\sqrt{2}$, corresponding to the structures where the upper wave lags (advances) the lower wave by a quarter of wavelength (as in Fig.~\ref{fig:fig5}(a,b)).


\subsection{Separatrix on a phase plane - classical ``fuzzy collapse''}
\label{subsec:4d}

\begin{figure}
   \centering -
    \includegraphics[width=1\textwidth]{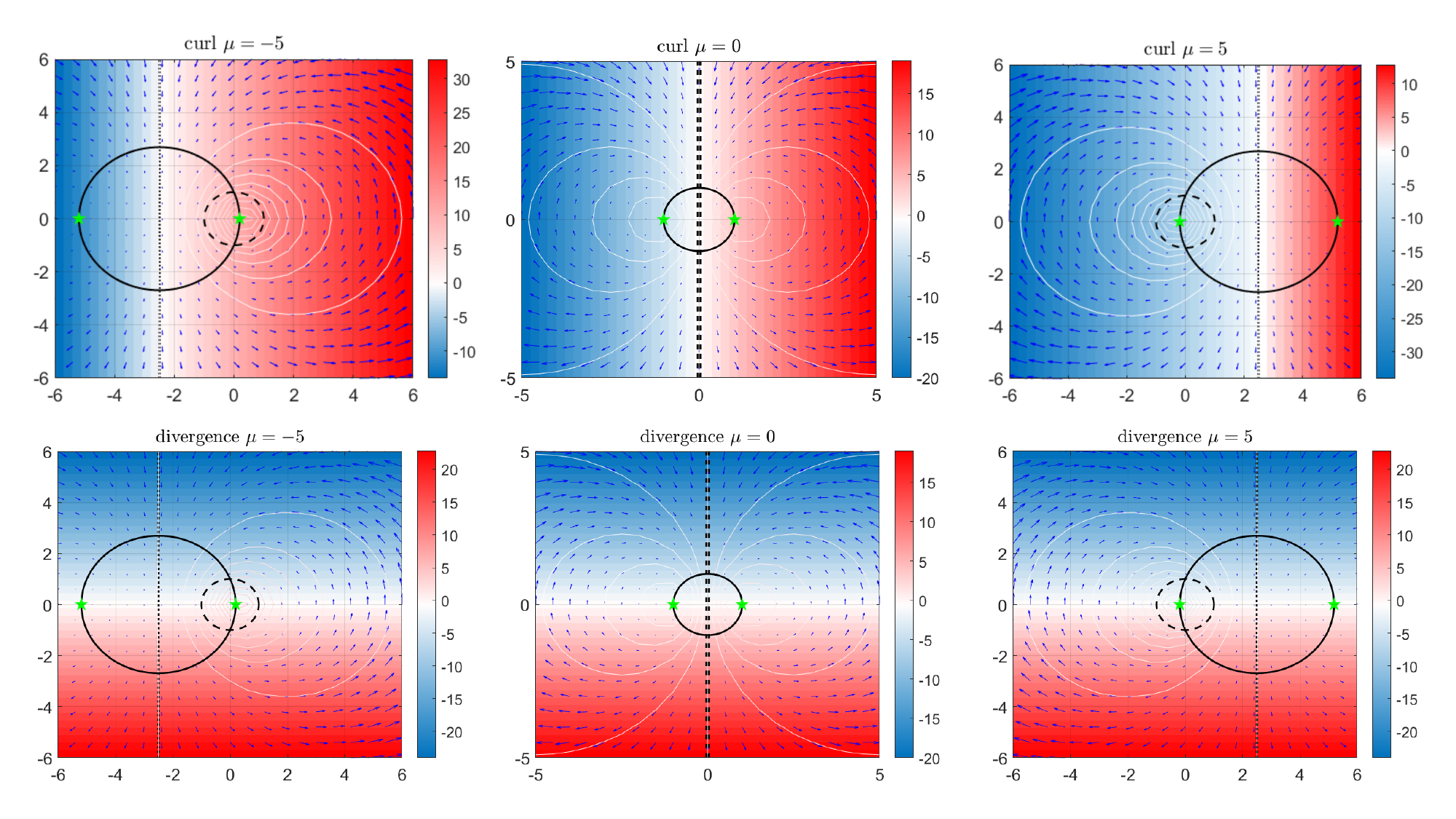}
   \caption{Phase plane flow of the dynamical system Eq.\,\eqref{UrUdelta}, for control parameter values $\mu = (-5, 0, 5)$. The fixed points Eq.\,\eqref{Fixedpoints} are indicated by the green stars. Colors show the values of the curl and the divergence fields in the upper and lower rows, respectively, according to Eq.\,\eqref{DivVor}. Unit circles around the origin are indicated by the dashed lines. Solid black circles of radius $\sqrt{\left({\mu \over 2}\right)^2 +1}$, connect the fixed points and are cut in half by the separatrix $x_s = \mu/2$. Trajectories on the phase plane encircle the fixed points, following Eq.\,\eqref{curves}, are indicated by the white closed curves. For the avoided crossing point, $\mu=0$, the dynamics on the separatrix for $\delta = \pm {\pi \over 2}$ corresponds respectively to the wave configurations of \ref{fig:fig5}(a,b).}
 \label{fig:fig7}
\end{figure}

Relate the wavefunction amplitude ratio to a radius: $r \equiv {\sqrt{\rho_1} \over \sqrt{\rho_2}} = {\sqrt{{\cal A}_1} \over \sqrt{{\cal A}_2}}$, equation set  \eqref{FeynModel3} can be reduced to the compact form of radial and azimuthal velocities on the polar coordinates $(r,\delta)$:
\begin{subequations}
\label{UrUdelta} 
\begin{align}
u_r & \equiv \dot{r} =-(r^2+1)\sin{\delta} \,  ,\\
u_{\delta} & \equiv r\dot{\delta} =(r^2 -1)\cos{\delta} - \mu r \, ,
\end{align}
\end{subequations}
where the time was scaled by the interaction coefficient, $t \mapsto \sigma t$. $\mu \equiv {\Delta \hat{\omega} / \sigma}$, indicating on the amount of coupling between the waves (the ratio between 
the frequency difference between the waves in isolation and their
interaction coefficient), can be regarded then as the control parameter of the dynamical system. 
In the frame of reference where $\overline{\hat{\omega}} = 0$, ${\mu / 2} = {\hat{\omega} / \sigma}$,  the two eigenstates Eq.\,\eqref{eigenstates} are the two neutral fixed points on the phase plane: 
\begin{subequations}
\label{Fixedpoints} 
\begin{align}
[r,\delta]^*_+ & =\left[\left({\mu \over 2} \pm \sqrt{\left({\mu \over 2}\right)^2 +1}\right),0 \right]\,  ,\\
[r,\delta]^*_- & =\left[\left(-{\mu \over 2} \pm \sqrt{\left({\mu \over 2}\right)^2 +1}\right),\pi \right]\, ,
\end{align}
\end{subequations}
where only positive values of $r^*$ are considered. 
The divergence and vorticity field of the phase plane flows become:
\begin{subequations}
\label{DivVor} 
\begin{align}
\nabla\cdot{\bf u} & ={1\over r}\left[\der{}{r}(r u_r)  + \der{u_{\delta}}{\delta}\right] = -4r\sin{\delta}\,  ,\\
\hat{\bf z}\cdot(\nabla\times{\bf u}) & ={1\over r}\left[\der{}{r}(r u_{\delta})  - \der{u_{r}}{\delta}\right] = 4r\cos{\delta} -2\mu\, ,
\end{align}
\end{subequations}
where the divergence vanishes on the neutral fixed points. A separatrix is obtained at $x_s={\mu \over 2}$, as evident when writing the velocity in the $x$ direction:
$u_x = u_r\cos{\delta} - u_{\delta}\sin{\delta} = y(\mu -2x)$.
Furthermore, from the conservation of charge/wave action and energy Eq.\,\eqref{Hamiltonian}, we find that the phase plane flow follows the closed curves, encircling the fixed points:
\be
\label{curves}
{(\mu r + 2 \cos{\delta})r\over r^2 +1} =  \mathrm{Const}\, . 
\ee
Examples of the phase plane flow are shown in Fig.~\ref{fig:fig7} for 
the control parameter values: $\mu = (-5,0,5)$.

The separatrix separates the left and the right sides of the phase plane into two ``regions of influence'' of the normal modes/eigenstates/fixed points. For any initial conditions of wave action amplitude ratio and phase difference, the dynamics satellites one of the fixed points, according to Eq.\,\eqref{curves}, without crossing to the other fixed point ``region of influence''. This is intriguing when considering the analog to the collapse of the wavefunction in quantum mechanics.
In the latter, the state of the system is obtained by the superposition of Eq.\,\eqref{generalsol}, but when a measurement is taken the system collapses immediately to one of the eigenstates ${\vec \Phi}_{\pm}$, with the respective  probability $A^2_{\pm}$, and then stays there forever, unless perturbed again. The collapse is in a sense exterior to the dynamics described by the  Schr\"{o}dinger equation and does not have a classical counterpart. Nevertheless, the phase-plane analysis suggests a classical ``fuzzy-collapse'' counterpart. Suppose that a wave maker generates a monochromatic wave perturbation with random amplitude ratio and phase difference between the two interfaces and then let it evolves according to Eq.\,\eqref{FeynModel3}. When we take a measurement, the dynamics is certainly not collapsing into one of the normal modes, but the measurement reveals at which side of the separatrix the system is, that is around which of the two eigenstates ${\vec \Phi}_{\pm}$ it keeps circulating.

\section{Discussion}
\label{sec:5}

This paper shows that the mathematical description of the interfacial Rossby wave interaction mechanism is equivalent to the description of Feynman's simplified model for the Josephson junction as a two-state coupled quantum system. 

The added value resulting from this comparison seems to be twofold. For the quantum system, it provides a simple mechanistic interpretation for its dynamics and eigenstates, especially for the central but obscured role of the phase difference between the macroscopic wavefunctions. For the fluid system, it sheds light on the mechanism of tunneling in terms of the superposition of the self and induced components of their cross-stream velocity fields.
Furthermore, the conservation laws of wave action and pseudoenergy obtain an interesting interpretation. Usually, the two Rossby wave action-at-a-distance interaction is considered to be a paradigm for barotropic and baroclinic instabilities \cite{hoskins1985use}. There the system is non-Hermitian, and while the dynamic equations can be represented elegantly by the canonical action-angle Hamilton equations \cite{heifetz2018generalized,heifetz2019normal}, the conserved quantities of the wave action and pseudoenergy must vanish to obtain modal instability - a somewhat confusing result. 
However here, for the stable interaction, the system is Hermitian and consists of a straightforward analog  to the elementary Planck–Einstein (particle) energy-(wave) frequency relation $E = \hbar \omega$. $\hbar$ is the elementary action unit of a single quantum particle, thus for the electron gas of density $\rho$, the energy density is $E = [\rho \hbar] \omega$. For Rossby waves the pseudoenergy $E = {\cal A}\, {\omega}$, where the wave action ${\cal A}$ is shown here to be equivalent to the electron gas density. Therefore, up to the scaling constant factors of the Planck number and the electron gas layers' volume, the conservation of wave action is analogous to the conservation of the total action of the Cooper pair quasi-particles. Furthermore, following the same logic, \eqref{Action0} can be regarded as a classical manifestation of the de Broglie (particle) momentum-wave number relation $P = \hbar k$.  
This suggests an intriguing classical perspective on wave-particle
duality.

One may ask why the analogy between such two remote physical systems, one quantum and the other classical, works. The correspondence is even more surprising when considering the fundamental dispersion relations of de Broglie matter waves and Rossby waves. The Schr\"{o}dinger equation for the wavefunction $\psi$ of a particle with mass $m$, in the absence of an external potential \cite{griffiths2018introduction}, and the vorticity equation for streamfunction $\phi$ on a negative constant vorticity gradient $ \beta$-plane, in the absence of mean flow \cite{vallis2017atmospheric}, read:
\be 
\label{Schrod-Rossby}
\dot{\psi} = i\left({\hbar \over 2m}\right) \nabla^2 \psi \, , \qquad
\dot{q} = \nabla^2 \dot{\phi} = i(\beta k)\phi\, .
\ee
Hence, for a given wavenumber $k$ in the $x$ 
direction the roles of the function and its Laplacian are flipped when we compare between the two equations. Consequently, the dispersion relations of plane waves of the form $e^{i({\bf k}\cdot{\bf x} - \omega t)}$ (where ${\bf k}$ and ${\bf x}$ are the wavenumber and position vectors) are very different: $\omega_{deBro} = {\hbar \over 2m}{\bf k}^2$ and $\omega_{Ros} = \beta{ k \over {\bf k}^2}$. The former can be interpreted as a sort of irrotational compressible quantum pressure waves \cite{heifetz2023broglie}, whereas the latter as vortical waves, resulted from advection of the mean flow vorticity. In the presence of a constant potential $U$ in the Schr\"{o}dinger equation, and a constant mean flow $\overline{u}$, in the vorticity equation,  Eq.\,\eqref{Schrod-Rossby} becomes:    
\begin{subequations}
\label{Elaborate-Schrod-Rossby} 
\begin{align}
\dot{\psi} = {i\left({\hbar \over 2m}\right) \nabla^2 \psi} -i {U\over \hbar}\psi \qquad \implies \qquad \omega_{deBro} = {\hbar \over 2m}{\bf k}^2 +{U\over \hbar}\, , \\
 \nabla^2 \dot{\phi} = i(\beta k)\phi-i(\overline{u} k)\nabla^2 {\phi}\, \qquad \implies \qquad \omega_{Ros} = k\left({\beta\over {\bf k}^2} + \overline{u} \right)\, .
\end{align}
\end{subequations}
Thus, for the de Broglie waves the constant potential adds energy but does not affect their group velocity, whereas for the Rossby waves the mean flow adds a Doppler shift.

In the Feynman model, the macroscopic wavefunction is assumed constant along each electron gas layer, thus $\nabla^2 \psi = 0$.
This assumption degenerates the de Broglie wave dynamics, consequently resulting in constant frequency $U/\hbar$ for each isolated gas layer. 
It also excludes the existence of the nonlinear quantum (Bohm) potential, appearing in the Madelung fluid-like representation of the Schr\"{o}dinger equation \cite{heifetz2015toward}.

For the interfacial Rossby waves at $y=y_0$, $\beta  = |\Delta \overline{q}_0|\delta(y-y_0)$, which makes the vorticity perturbation a delta function at the interface with a wavy streamfunction in the $x$ direction and evanescent structure in the $y$ direction. For $\overline{u}_0 <0$, this gives the dispersion relation of Eq.\,\eqref{FrequencySingle}  $\left({1\over 2}|\Delta \overline{q}_0| -k|\overline{u}_0|\right)$. Consequently, $U/\hbar \Longleftrightarrow \left({1\over 2}|\Delta \overline{q}_0| -k|\overline{u}_0|\right)$, thus for given values of the interfacial mean flow and vorticity gradient, a change in the wavenumber $k$ accounts for a change in the ground state $U$ of the Cooper pairs in the electron gas layer. 
The other ingredient that fits in is the simple representation for the tunneling between the superconducting layers in the Feynman model ($\sigma$) that is represented by the evanescent structure of the cross-stream velocity induced by each interfacial waves $\sigma  \mapsto \sqrt{|\Delta\overline{q}_1||\Delta\overline{q}_2|}\, {e^{-kY}/ 2}\, $. 

The equivalence between the two systems has therefore resulted from the gross simplification of the physics describing the two systems. 
It would be interesting to examine whether such equivalence is maintained when considering more realistic setups. For instance, a straightforward extension could be a case where the superconducting layers are not entirely homogeneous, so that the gradients of the phases yield currents within the layers. In the fluid system this may be equivalent to a slow variation of the mean flow and the mean vorticity gradients across the interfaces, so that the Rossby wave dynamics would be described using WKB approximation. We leave this analysis for a future work.

The comparison between the systems raises the question whether the two-Rossby wave system can serve as a sort of a hydrodynamic qubit device (which would obviously be enormous in size and extremely slow in comparison to trapped ions or superconducting quantum interference devices (SQUIDs) \cite{kockum2019quantum}), even in principle. However, despite of the equivalence between the two systems, including the ``fuzzy collapse'' behavior, the fluid system is classical, thus lacks the features of quantum collapse and entanglement, which are essential requirements to perform quantum computation.



\vspace{1cm}
{\noindent\large \bf Acknowledgments}\\
We thank the two referees for their supportive positive comments and  useful suggestions.
 Eyal Heifetz is grateful as well to Erik Gengel, Noa Feldman and Yuval Dagan for fruitful discussions.

\appendix

\section{Action-at-a-distance as tunneling}
\label{Appendix-A}

\setcounter{equation}{0} \renewcommand{\theequation}{A\arabic{equation}} 

\begin{figure}
   \centering -
    \includegraphics[width=1\textwidth]{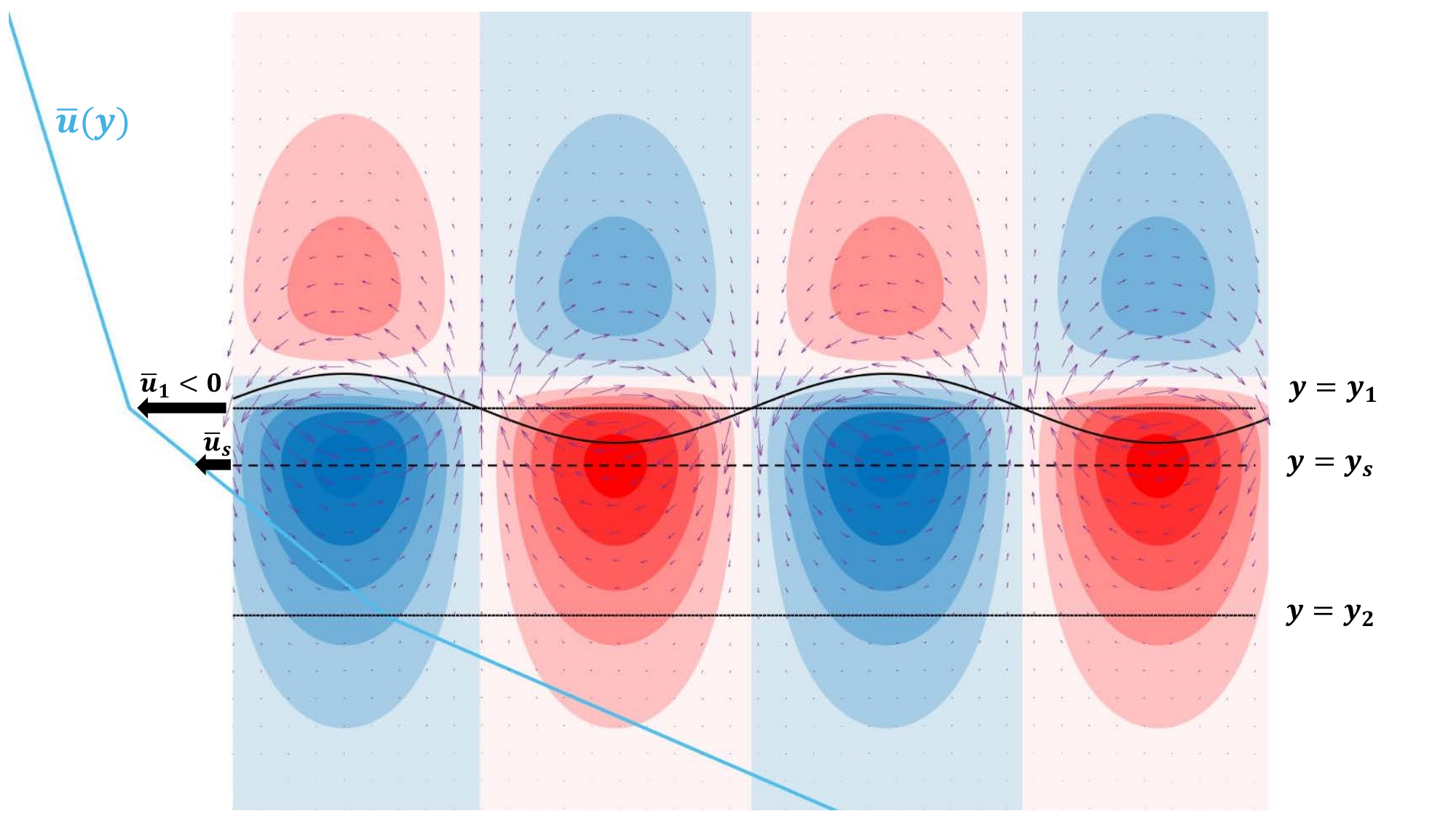}
   \caption{The velocity and pressure fields induced by the interfacial wave, positioned at $y=y_1$. The undulated solid line indicates the wave displacement $\eta$ at the interface. The velocity arrows indicates the Rossby propagation mechanism, translating the displacement to the right counter the mean flow ${\overline{u}}_1$. 
   The velocity field decays exponentially away from the interface but induces a non zero field on interface $y_2$ (the tunneling effect). The induced pressure filed Eq.\,\eqref{A:pressure}, is indicated by the colored contours where red (blue) refers to positive (negative) values. At the steering level $y_s$, $\hat{c}_1  = {\overline{u}}_1(y_s)$. This is enabled as in the $x$ direction the pressure gradient force balances the acceleration/deceleration of $u_1(y_s)$, resulting from the vertical advection of the mean flow by the velocity perturbation $v_1(y_s)$. In the $y$ direction the pressure gradient must vanish there in order to allow the $v$ field to be passively advected by ${\overline{u}}_1(y_s)$.}
 \label{fig:fig8}
\end{figure}

As discussed in subsection \ref{subsec:4a}, the analog for tunneling in the fluid system is the action-at-a-distance between the waves, mediated by the far field cross-stream velocity $v^i$, induced by each wave on the other.   
For a classical incompressible fluid, action-at-a-distance is not ``spooky'' as in quantum mechanics \cite{bell2004speakable}, in the sense that the speed of sound (playing an equivalent role to the speed of light in quantum mechanics), is assumed infinite (zero Mach number), thus information is assumed to be travelling ``infinitely'' fast within the fluid. 

The constant shear layer in between the two interfaces is the analog for the insulator in the Josephson junction device. As the vorticity perturbation field is non zero only at the interfaces, the in between shear layer is transparent to it. However, one may ask how the two components of the induced velocity field, together with the induced pressure field, resist the shear and maintain untilted structures.

Consider then profile Eq.\,\eqref{TwoShear} and the fields induced by the upper interfacial wave, $q_1 =\hat{Q}_1\delta(y-y_1)e^{ik(x- \hat{c}_1 t)}$ on the shear layer below  $y_2<y<y_1$. Applying Eq.\,\eqref{SFKSingle} for $y_0 \mapsto y_1$, the induced fields satisfy: 
$(\phi_1, u_1, v_1) = 0.5(-k^{-1}, 1, -i){\hat{Q}_1}e^{k(y-y_1)}e^{ik(x- \hat{c}_1 t)}$, together with a pressure field $p_1 = \tilde{p}(y)e^{ik(x- \hat{c}_1 t)}$, that is yet need to be found. 
The steering level, that is the height $y_s$, where the mean flow is equal to the  phase velocity of the wave: $\overline{u}(y_s) = \hat{c}_1 = (- |\overline{u}_1| + {|\Delta \overline{q}_1|\over 2k})$, is located at  $y_s = y_1 - {1\over 2k}{|\Delta \overline{q}_1|\over \overline{q}_M}$, where we choose ${|\Delta \overline{q}_1|\over \overline{q}_M} = 1 - {\overline{q}_T\over \overline{q}_M}< 2kY$, to ensure the steering level to be found inside the middle shear layer. Above the steering level $\hat{c}_1 > \overline{u}(y)$, whereas below it $\hat{c}_1 < \overline{u}(y)$. Hence, as indicated from the momentum equation of Eq.\,\eqref{EQsLinFlow}, to maintain the untilted structure of the induced fields outside the steering level, the pressure gradient force implied by the wave field, together with the cross-stream advection of the mean shear flow, must balance the advection of the mean flow. Substitute in the two-component of the momentum equations we obtain:
\be
\label{A:momentum}
[\hat{c}_1  - \overline{u}(y)]u_1 = p_1 - \overline{q}_M \phi_1\, ; \qquad
[\hat{c}_1  - \overline{u}(y)]v_1 =
-{i \over k}\der{p}{y}\, ,
\ee
from which we obtain the induced pressure field:
\be
\label{A:pressure}
p_1 = -0.5\left[{\overline{q}_T + \overline{q}_M \over 2k} + \overline{q}_M(y_1 - y)\right]{\hat{Q}_1}e^{k(y-y_1)}e^{ik(x- \hat{c}_1 t)} .
\ee
Thus, the wave's induced pressure field is in anti-phase with its vorticity field (fig.~\ref{fig:fig8}). At its home base interface ($y=y_1$),
$p_1^1 = -\left({\overline{q}_T + \overline{q}_M \over 4k}\right){\hat{Q}_1}e^{ik(x- \hat{c}_1 t)} = \left({\overline{q}_T + \overline{q}_M \over 2}\right)\phi_1^1 \equiv \overline{q}^1 \phi_1^1$ \footnote{This stands in agreement with the limit of the stream-wise momentum equation of Eq.\,\eqref{A:inducedfields} at the interface.  $u_1^1$ is evaluated zero, as $u_1$ flips sign when crossing $y_1$, and the mean vorticity $\overline{q}^1$ is evaluated as the averaged mean vorticity at its two sides}. 
At the steering level $p^{s}_1 = \overline{q}_M \phi^{s}_1$ and $\left(\der{p_1}{y}\right)_{y_s} = 0$.
The complete untilted wave structure that is induced (``tunneled'') from the interfacial vorticity Rossby wave at interface $1$ onto interface $2$ is:  
\be
\label{A:inducedfields}
(u, v, p)_1^2 = 
0.5\left[1, -i, -\left({\overline{q}_T + \overline{q}_M \over 2k} + \overline{q}_M Y\right)\right]{\hat{Q}_1}e^{-kY}e^{ik(x- \hat{c}_1 t)}.
\ee

\section{Equivalence between the Hamiltonians of the Rossby wave and the Josephson junction systems}
\label{Appendix-B}

\setcounter{equation}{0} \renewcommand{\theequation}{B\arabic{equation} 

We first write the Josephson junction Hamiltonian \eqref{Hamiltonian} in terms of the Rossby wave system properties, implementing 
$ \rho_{1,2} \mapsto {\cal A}_{1,2}\,$: 
\begin{equation}
\label{B:Rossby_Hamiltonian1}
H =  
\sum_{i=1}^2 {\cal A}_i{\hat \omega}_{i}  + 2\sigma\sqrt{{\cal A}_1{\cal A}_2}\cos{(\theta_1 - \theta_2)}\, .
\end{equation}
Next we substitute \eqref{omega12Rossby} and \eqref{Action12} in \eqref{B:Rossby_Hamiltonian1} and use the Green function expressions $(G^s, G^i) =
-(1, e^{-kY})/2k\,$ and the scaled interaction coefficient $\sigma  = \sqrt{|\Delta\overline{q}_1||\Delta\overline{q}_2|}\, {e^{-kY}/ 2}\,$, to obtain:
\begin{equation}
\label{B:Rossby_Hamiltonian2}
H ={1\over 4}\left[{\hat{Q}_1^2 \over 2k} + {\hat{Q}_2^2 \over 2k}  +
{e^{-k|Y|}\over k}\hat{Q}_1 \hat{Q}_2\cos{(\theta_1 - \theta_2)} -
\left( {|\overline{u}_1| \over |\Delta\overline{q}_1|}\hat{Q}_1^2 + 
{|\overline{u}_2| \over |\Delta\overline{q}_2|}\hat{Q}_2^2
\right)\right]\, .  
\end{equation}
It is now left to show that \eqref{B:Rossby_Hamiltonian2} is obtained as well  when we substitute \eqref{TwoVortJumps} and $Q(y) = \hat{Q}_1\delta(y-y_1) + \hat{Q}_2\delta(y-y_2)$ in \eqref{ExplicitPE}.
The inner integral in \eqref{ExplicitPE} over the dummy variable $y'$ gives:
$$
\int_{-L_y/2}^{L_y/2} Q(y')\cos{(\theta(y) -\theta(y'))}{e^{-k|y-y'|}\over 2k}dy' = 
{1\over 2k}\left[\hat{Q}_1 \cos{(\theta(y) - \theta_1)}e^{-k|y-y_1|} + \hat{Q}_2 \cos{(\theta(y) - \theta_2)}e^{-k|y-y_2|}  
\right].
$$
Consequently:
$$
\int_{-L_y/2}^{L_y/2}{Q(y)\over 4}\left[
\int_{-L_y/2}^{L_y/2} Q(y')\cos{(\theta(y) -\theta(y'))}    {e^{-k|y-y'|}\over 2k}dy' \right] dy = {1\over 4}\left[{\hat{Q}_1^2 \over 2k} + {\hat{Q}_2^2 \over 2k}  +
{e^{-k|Y|}\over k}\hat{Q}_1 \hat{Q}_2\cos{(\theta_1 - \theta_2)}\right].
$$
The last term of \eqref{ExplicitPE} becomes:
$$-\int_{-L_y/2}^{L_y/2}{Q(y)\over 4}{\overline{u}(y)\, Q (y) \over {\overline{q}}_{y}(y)}  dy=
-{1\over 4}
\left( {|\overline{u}_1| \over |\Delta\overline{q}_1|}\hat{Q}_1^2 + 
{|\overline{u}_2| \over |\Delta\overline{q}_2|}\hat{Q}_2^2\right).
$$
Gathering now the two RHS terms in the last two equations we obtain the RHS of \eqref{B:Rossby_Hamiltonian2}, so that indeed $E_{Rossby} \leftrightarrow  H_{Josephson}$. 

We note that \eqref{B:Rossby_Hamiltonian2} indicates that the Fj\o rtoft stability condition is not satisfied as $\mathrm{sgn}({\overline u}_i) = \mathrm{sgn}({{\overline{q}}_{y}}_i)$ for each of the two interfaces ($i=1,2$). Nevertheless, the Rayleigh stability condition is satisfied as $\mathrm{sgn}({{\overline{q}}_{y}}_1) = \mathrm{sgn}({{\overline{q}}_{y}}_2)$ (Fig.~\ref{fig:fig4}). Thus, for this setup, it is the conservation of the wave action ${\cal A}$, rather than the conservation of energy $H$, that assures modal stability.

}
\bibliography{apssamp}

\end{document}